\pdfoutput=1

\def\full{0}
\def\authnotes{1}

\ifnum\full=0
\documentclass[letterpaper,twocolumn,10pt]{article}
\usepackage{usenix}
\usepackage[letterpaper,hmargin=0.85in,vmargin=0.95in]{geometry}

\else
\documentclass[letterpaper,10pt]{article}
\fi

\usepackage{epsfig,endnotes}

\usepackage[hyphens]{url}
\usepackage[usenames,dvipsnames]{color}
\usepackage{tabularx}
\usepackage{amsfonts}
\usepackage{setspace}
\usepackage{arydshln}
\usepackage{stmaryrd}
\usepackage{cite}

\usepackage{verbatimbox}

\usepackage{subfigure}

\usepackage{authblk}

\usepackage[font={small}]{caption}

    \newtheorem{thm}{\hspace*{-1em}Theorem}[section]
    \newtheorem{lem}[thm]{\hspace*{-1em}Lemma}
    \newtheorem{cor}[thm]{\hspace*{-1em}Corollary}
    \newtheorem{propo}[thm]{\hspace*{-1em}Proposition}
    \newtheorem{fct}[thm]{\hspace*{-1em}Fact}
    \newtheorem{defn}[thm]{\hspace*{-1em}Definition}
    \newtheorem{exmp}[thm]{\hspace*{-1em}Example}
    \newtheorem{assm}[thm]{\hspace*{-1em}Assumption}
    \newtheorem{clm}[thm]{\hspace*{-1em}Claim}
    \newtheorem{techclm}[thm]{Technical Claim}
    \newtheorem{rem}[thm]{Remark}
    \newtheorem{cons}[thm]{Construction}

\newcommand{\thmvspace}{\vspace*{.5em}}
    {\end{rm}\end{thm}\thmvspace}
    {\end{rm}\end{lem}\thmvspace}
    {\end{rm}\end{cor}\thmvspace}
    {\end{rm}\end{propo}\thmvspace}
    {\end{rm}\end{fct}\thmvspace}
    {\end{em}\end{defn}\thmvspace}
    {\end{em}\end{exmp}\thmvspace}
    {\end{em}\end{assm}\thmvspace}
    {\end{rm}\end{clm}\thmvspace}
    {\end{rm}\end{techclm}\thmvspace}
    {\end{em}\end{rem}\thmvspace}
    {\end{em}\end{cons}\thmvspace}

    \newcount\proofqeded
    \newcount\proofended
    \def\qed{
    \end{rm}\addtolength{\parskip}{-0pt}
    \setlength{\parindent}{\saveparindent}
    \global\advance\proofqeded by 1 }
     {\proofstart}%
     {\ifnum\proofqeded=\proofended\qed\fi \global\advance\proofended by 1
      \medskip}
    \makeatletter
    \def\proofstart{\@ifnextchar[{\@oprf}{\@nprf}}
    \def\@oprf[#1]{\begin{rm}\protect\vspace{6pt}\noindent{\bf Proof of #1:\
    }%
    \addtolength{\parskip}{5pt}\setlength{\parindent}{0pt}}
    \def\@nprf{\begin{rm}\protect\vspace{6pt}\noindent{\bf Proof:\ }%
    \addtolength{\parskip}{5pt}\setlength{\parindent}{0pt}}
    \makeatother

\newlength{\saveparindent}
\newlength{\saveparskip}
\DeclareMathAlphabet{\mathsl}{OT1}{cmr}{m}{sl}
\DeclareMathAlphabet{\mathsc}{OT1}{cmr}{m}{sc}

\newcommand{\secref}[1]{\mbox{Section~\ref{#1}}}
\newcommand{\syref}[1]{\mbox{\S~\ref{#1}}}

\newcommand{\figref}[1]{\mbox{Figure~\ref{#1}}}

\newcommand{\SetFigFont}[5]{} 
\newcommand{\SetFigFontNFSS}[5]{} 

\newcounter{ctr}

\newcounter{ectr}

\newenvironment{newitemize}{%
\begin{list}{\mbox{}\hspace{5pt}$\bullet$\hfill}{\labelwidth=15pt%
\labelsep=5pt \leftmargin=20pt \topsep=3pt%
\setlength{\listparindent}{\saveparindent}%
\setlength{\parsep}{\saveparskip}%
\setlength{\itemsep}{3pt} }}{\end{list}}

\newenvironment{centerlist}{%
\begin{list}{\mbox{}}{\labelwidth=0pt%
\labelsep=0pt \leftmargin=0pt \topsep=10pt%
\setlength{\listparindent}{\saveparindent}%
\setlength{\parsep}{\saveparskip}%
\setlength{\itemsep}{10pt} }}{\end{list}}

\newlength{\savejot}
\newenvironment{newmath}{\begin{displaymath}%
\setlength{\abovedisplayskip}{4pt}%
\setlength{\belowdisplayskip}{4pt}%
\setlength{\abovedisplayshortskip}{6pt}%
\setlength{\belowdisplayshortskip}{6pt} }{\end{displaymath}}

\newenvironment{newequation}{\begin{equation}%
\setlength{\abovedisplayskip}{4pt}%
\setlength{\belowdisplayskip}{4pt}%
\setlength{\abovedisplayshortskip}{6pt}%
\setlength{\belowdisplayshortskip}{6pt} }{\end{equation}}

\renewcommand{\Sigma}{\{0,1\}}

\newcommand{\smidge}{{\kern .05em}}

\newcolumntype{y}[1]{%
>{\hspace{0pt}}p{#1}}%

\newcolumntype{x}[1]{%
>{\centering\hspace{0pt}}p{#1}}%

\newcommand{\verylongrightarrow}[1]             
      {\setlength{\unitlength}{.01in}           
      \begin{picture}(#1,1) \put(0,0){\vector(1,0){#1}} \end{picture}}
\newcommand{\verylongbotharrow}[2]             
      {\setlength{\unitlength}{.01in}           
      \begin{picture}(#2,1) \put(#1,0){\vector(1,0){#1}}
                            \put(#1,0){\vector(-1,0){#1}} \end{picture}}

\renewcommand{\choose}[2]{{{#1}\atopwithdelims(){#2}}}

\newcommand{\E}{{\mbox{\bf E}}}

\newcommand{\Prob}[1]{\Pr\left[\: #1 \:\right]}

\makeatletter

\def\subsubsection{\@startsection{subsubsection}{3}{\z@}{-2.25ex plus
 -1ex minus -.2ex}{1.5ex plus .2ex}{\sf}}

\newcommand{\email}[1]{\texttt{#1}}

\newcommand{\authnote}[2]{\ifnum\authnotes=1\begin{quote}\textbf{#1 says:} #2\end{quote}\fi}
\newcommand{\authfnote}[2]{\ifnum\authnotes=1\footnote{\textbf{#1 says:} #2}\fi}

\renewcommand{\paragraph}[1]{\vspace{.6em}\noindent\textbf{#1}\hspace*{.5em}}
\newcommand{\iparagraph}[1]{\vspace{.6em}\noindent\textit{#1}\hspace*{.5em}}

\newcounter{mynote}[section]
\newcommand{\notecolor}{blue}
\newcommand{\tsnotecolor}{NavyBlue}
\newcommand{\thenote}{\thesection.\arabic{mynote}}
\newcommand{\tnote}[1]{\ifnum\authnotes=1\refstepcounter{mynote}{\bf \textcolor{\notecolor}{$\ll$TCR~\thenote: {\sf #1}$\gg$}}\fi}
\newcommand{\tsnote}[1]{\ifnum\authnotes=1\refstepcounter{mynote}{\bf \textcolor{\tsnotecolor}{$\ll$TS~\thenote: {\sf #1}$\gg$}}\fi}
\newcommand{\vnote}[1]{\ifnum\authnotes=1\refstepcounter{mynote}{\bf \textcolor{\notecolor}{$\ll$VV~\thenote: {\sf #1}$\gg$}}\fi}
\newcommand{\fixme}[1]{\ifnum\authnotes=1\textbf{\textcolor{red}{[FIXME: #1]}}\fi}

\newcommand{\ignore}[1]{}

\newcommand{\bnm}{\begin{newmath}}
\newcommand{\enm}{\end{newmath}}
\newcommand{\bne}{\begin{newequation}}
\newcommand{\ene}{\end{newequation}}

\newcommand{\naive}{na\"{i}ve}

\def\ourtitle{A Placement Vulnerability Study in Multi-Tenant Public Clouds}

\def\ourkeywords{\textbf{Keywords}: co-location detection, multi-tenancy, cloud security}

\makeatletter
\renewcommand\AB@affilsepx{, \protect\Affilfont}
\makeatother

\def\@fnsymbol#1{\ensuremath{\ifcase#1\or \mathsection\or *\or \mathparagraph\or \dagger\or \ddagger\or
\|\or **\or \dagger\dagger  \or \ddagger\ddagger \else\@ctrerr\fi}}

\title{\Large \textbf{\ourtitle}\thanks{This is the full version of an earlier paper published at USENIX Security 2015~\cite{venkat-placement15}.}
}

\author[$\star$]{Venkatanathan Varadarajan}
\author[$\ddag$]{Yinqian Zhang}
\author[ ]{Thomas Ristenpart\thanks{Work primarily done while at
the University of Wisconsin-Madison.}}
\author[$\star$]{Michael Swift}
\affil[$\star$]{\textit{University of Wisconsin-Madison}}
\affil[$\ddag$]{\textit{The Ohio State University}}
\affil[$\dag$]{\textit{Cornell Tech}}
\affil[
]{\normalsize\email{\{venkatv,swift\}@cs.wisc.edu}, \email{yinqian@cse.ohio-state.edu},\email{
ristenpart@cornell.edu}}
\date{}

\begin{document}

\maketitle

\begin{abstract}

\noindent Public infrastructure-as-a-service clouds, such as Amazon
EC2, Google Compute Engine (GCE) and Microsoft Azure allow clients to run
virtual machines (VMs) on shared physical infrastructure.  This
practice of multi-tenancy brings economies of scale, but also
introduces the risk of sharing a physical server with an arbitrary and potentially
malicious VM. Past works have demonstrated how to place a VM alongside
a target victim (co-location) in early-generation clouds and how to
extract secret information via side-channels. Although there have been
numerous works on side-channel attacks, there have been no studies
on placement vulnerabilities in public clouds since the adoption of
stronger isolation technologies such as Virtual Private Clouds (VPCs).

We investigate this problem of placement vulnerabilities and
quantitatively evaluate three popular public clouds for their
susceptibility to co-location attacks. We find that adoption of
new technologies (e.g., VPC) makes many prior attacks, such as cloud
cartography, ineffective. We find new ways to reliably test for
co-location across Amazon EC2, Google GCE, and Microsoft Azure. We
also find ways to detect co-location with victim web servers in a multi-tiered
cloud application located behind a load balancer.

We use our new co-residence tests and multiple customer accounts to launch VM instances
under different strategies that seek to maximize the likelihood of
co-residency. We find that it is much easier (10$\times$ higher success rate)
and cheaper (up to \$114 less) to achieve co-location in these three clouds
when compared to a secure reference placement policy.

\\

\noindent \ourkeywords
\end{abstract}


\ifnum\full=1
\newpage
\tableofcontents
\newpage
\fi
\pagestyle{plain}

\section{Introduction}
\label{sec:intro}


Public cloud computing offers easy
access to relatively cheap compute and storage resources. Cloud providers are
able to sustain this cost-effective solution through {\em
  multi-tenancy}, where the infrastructure is shared between
computations run by arbitrary customers over the Internet. This
increases utilization compared to dedicated infrastructure, allowing
lower prices.

However, this practice of multi-tenancy also enables various security
attacks in the public cloud.  Should an adversary be able to launch
a virtual machine on the same physical host as a victim, making the two VMs
co-resident (sometimes the term co-located is used),
there exist
attacks that break the logical isolation provided by virtualization to
breach
confidentiality~\cite{zhang12,yarom14flush,zhang2014cross,xu2011exploration,rist:hey-you:ccs:2009,wu2012whispers}
or degrade the performance~\cite{venkat:rfa:ccs2012,xen-cycle-stealing}
of the victim. Perhaps most notable are the side-channel attacks that
steal private keys across the virtual-machine isolation boundary
by cleverly monitoring shared resource
usage~\cite{zhang12,yarom14flush,zhang2014cross}.

Less understood is the ability of adversaries to arrange for 
co-residency in the first place. In general, doing so consists of using a {\em launch strategy}
together with a mechanism for {\em co-residency detection}.
The only prior work on obtaining co-residency~\cite{rist:hey-you:ccs:2009}  
showed simple network-topology-based co-residency checks along with low-cost launch strategies that
obtain a high probability of achieving co-residency compared to simply launching
as many VM instances as possible. 
When such advantageous strategies exist, we
say the cloud suffers from a placement vulnerability. 
Since then, Amazon has
made several changes to their architecture, including removing the ability to do
the simplest co-residency check. Whether placement vulnerabilities exist in
other public clouds has, to the best of our knowledge, never been explored.



In this work, we provide a framework to systematically evaluate public
clouds for placement vulnerabilities and show that three popular
public cloud providers may be vulnerable to co-location attacks.
More specifically, we set out to answer four questions:
\begin{newitemize} 
 \item Can co-residency be effectively detected in modern public clouds?
 \item Are known launch strategies~\cite{rist:hey-you:ccs:2009} still effective
 in modern clouds? 
 \item Are there any new exploitable placement vulnerabilities?
 \item Can we quantify the money and time required of an adversary to achieve a certain
 probability of success?  
\end{newitemize}


We start by exploring the efficacy of prior co-residency tests
(\syref{sec:cores-test}) and develop more reliable tests for our placement
study (\syref{ssec:corestest}). We also find a novel test to detect co-residency
with VMs uncontrolled by the attacker by just using their public interface even when they are
behind a load balancer (\syref{ssec:uncoop-test}).

We use multiple customer accounts across three popular cloud providers, launch
VM instances under different scenarios that may affect the placement algorithm,
and test for co-residency between all launched instances. We analyze three
popular cloud providers, Amazon Elastic Compute Cloud (EC2)~\cite{amazon:ec2}, Google
Compute Engine (GCE)~\cite{gce} and Microsoft Azure (Azure)~\cite{azure}, for
vulnerabilities in their placement algorithm. After exhaustive experimentation
with each of these cloud providers and at least 190 runs per cloud provider, we
show that an attacker can still successfully arrange for co-location 
(\syref{sec:results}). We find new launch strategies in these three clouds 
that obtain co-location faster (10x higher success rate)
and cheaper (up to \$114 less)
when compared to a secure reference placement policy. 

Next, we start by giving some background on public clouds
(\syref{sec:background}) and then define our threat model
(\syref{sec:threat-model}). We conclude the paper with related and future work
(\syref{sec:relwork} and \syref{sec:conclude}, respectively).

\section{Background}
\label{sec:background}

\vspace{-0.1in}
\paragraph{Public clouds.}
Infrastructure-as-a-service (IaaS) public clouds, such as Amazon EC2,
Google Compute Engine and Microsoft Azure, provide a management
interface for customers to launch and terminate VM
instances with a user-specified configuration. Typically, users
register with the cloud provider for an account and use the cloud
interface to specify VM configuration, which includes instance type,
disk image, data center or region to host the VMs, and then launch VM
instances. In addition, public clouds also provide many higher-level
services that monitor load and automatically launch or terminate
instances based on the workload~\cite{rightscale,
aws-elastic-bs, gce-autoscaler}. These services internally use the
same mechanisms as users to configure, launch and terminate VMs.

The provider's VM launch service receives from a client a desired set
of parameters describing the configuration of the VM. The service then
allocates resources for the new VM; this process is called {\em VM
provisioning}. We are most interested in the portion of VM
provisioning that selects the physical host to run a VM, which we call
the {\em VM placement algorithms}. The resulting VM-to-host mapping we
call the {\em VM placement}.  The placement for a specific virtual
machine may depend on many factors: the load on each machine, the
number of machines in the data center, the number of concurrent VM
launch requests, etc.


\begin{figure}
\centering\
\footnotesize
\begin{tabular}{|l|l|l|}
  \hline
  \textbf{Type} & \textbf{Variable}\\
  \hline
  & \# of customers \\
  & \# of instances launched per customer \\
  Placement & Instance type \\
  Parameters & Data Center (DC) or Region \\  
  & Time launched \\
  & Cloud provider \\
  \hline
  & Time of the day \\
  Environment & Days of the week\\
  Variable & Number of in-use VMs\\
  & Number of machines in DC\\
  \hline
\end{tabular}
\caption{\textbf{List of placement variables.}}
\label{fig:variables}
\end{figure}
While cloud providers do not generally publish their VM placement algorithms,
there are several variables under the control of the user that could affect the VM
placement, such as time-of-day, requested data center, and number of
instances. A list of some notable parameters are given in
\figref{fig:variables}. By controlling these variables, an adversary can
partially influence the placement of VMs on physical machines that may also host
a target set of VMs. We call these variables {\em placement variables} and the
set of values for these variables form a {\em launch strategy}. An example
launch strategy is to launch 20 instances 10 minutes after triggering an
auto-scale event on a victim application. This is, in fact, a launch strategy
suggested by prior work~\cite{rist:hey-you:ccs:2009}.


\paragraph{Placement policies.} 
VM placement algorithms used in public clouds often aim to increase
data center efficiency, quality of service, or both by realizing some
\emph{placement policy}. For instance, a policy that aims to increase
data center utilization may pack launched VMs on a single machine.
Similarly policies that optimize the time to provision a VM, which
involves fetching an image over the network to the physical machine and
booting, may choose the last machine that used the same VM image, as
it may already have the VM image cached on local disks. Policies may
vary across cloud providers, and even within a provider.


Public cloud placement policies, although undocumented, often exhibit
behavior that is externally observable. One example is {\em parallel
placement locality}~\cite{rist:hey-you:ccs:2009}, in which VMs
launched from different accounts within a short time window are
often placed on the same physical machine. Two instances launched
sequentially, where the first instance is terminated before the launch
of the second one, are often placed on the same physical machine, a
phenomenon called \emph{sequential placement locality}~\cite{rist:hey-you:ccs:2009}. 


These placement behaviors are artifacts of the two placement policies
described earlier, respectively. Other examples of policies and
resulting behaviors exist as well. VMs launched from the same accounts
may either be packed on the same physical machine to maximize locality
(and hence co-resident with themselves) or striped across
different physical machines to maximize redundancy (and hence
never co-resident with themselves).  In the course of normal usage,
such behaviors are unlikely to be noticed, but they can be measured
with careful experiments.


\paragraph{Launch strategies.} 
An adversary can exploit placement behaviors to increase the
likelihood of co-locating with target victims. As pointed out by
Ristenpart et al.~\cite{rist:hey-you:ccs:2009}, parallel placement
locality can be exploited by triggering a scale-up event on target
victim by increasing their load, which will cause more victim VMs to
launch. The adversary can then simultaneously (or after a time lag)
launch multiple VMs some of which may be co-located with the newly
launched victim VM(s).


In this study, we develop a framework to systematically evaluate
public clouds against launch strategies and uncover previously
unknown placement behaviors. We approach this study by (i) identifying
a set of placement variables that characterize a VM, (ii) enumerating
the most interesting values for these variables, and (iii) quantifying
the cost of such a strategy, if it in fact exposes a co-residency
vulnerability. We repeat this for three major public cloud providers:
Amazon EC2, Google Compute Engine, and Microsoft Azure. Note that the
goal of this study is not to {\em reverse engineer} the exact details of the
placement policies,  but rather to identify launch
strategies that can be exploited by an adversary.

\paragraph{Co-residency detection.}
A key technique for understanding placement vulnerabilities is
detecting when VMs are co-resident on the same physical machine (also
termed {\em co-locate}). Ristenpart et al.~\cite{rist:hey-you:ccs:2009} 
proposed several co-residency
detection techniques and used them to identify several placement
vulnerabilities in Amazon EC2. As co-resident status
is not reported directly by the cloud provider, these detection
methods are usually referred to as side-channel based techniques,
which can be further classified into two categories: {\em logical
side-channels} or {\em performance side-channels}.


 
\iparagraph{Logical side-channels:} Logical side-channels allow information
leakage via logical resources that are observable to a software program, e.g.,
IP addresses, timestamp counter values.  Particularly in Amazon EC2, each VM is
assigned two IP addresses, a \emph{public} IP address for communication over the
Internet, and a private or \emph{internal} IP address for intra-datacenter
communications. The EC2 cloud infrastructure allowed translation of public IP
addresses to their internal counterparts. This translation revealed the topology
of the internal data center network, which allowed a remote adversary to map the
entire public cloud infrastructure and determine, for example, the availability
zone and instance type of a victim. Furthermore, co-resident VMs tended to have
adjacent internal IP addresses. 


Logical side-channels can also be established
via shared timestamp counters. In prior work, skew in timestamp
counters were used to fingerprint a physical
machine~\cite{kohno2005remote}, although this technique has not yet
been explored for co-residency detection. Co-residency detection can possibly
be performed via any shared
software state between the two customers. In the context of container-based
platform-as-a-service (PaaS) clouds, where customers share the same 
operating system, example of logical side-channels include interrupt 
counts and process statistics reported in \texttt{procfs}. 


\iparagraph{Performance side-channels:} Performance side-channels are created
when performance variations due to resource contention are observable.  Such
variations can be used as an indicator of co-residency. For instance, network
performance has been used for detecting
co-residence~\cite{rist:hey-you:ccs:2009,venkat:rfa:ccs2012}.  This is because
hypervisors often directly relay network traffic between VMs on the same host,
providing detectably shorter round-trip times than between VMs on different
hosts.  

Covert channels, as a special case of side-channels, can be established between
two VMs that are cooperating in order to detect co-residency. For purposes of
co-residency detection, covert channels based on shared hardware resources, such
as last level caches (LLCs) or local storage disks, can be exploited by one VM
to detect performance degradation caused by a co-resident VM. Covert channel
detection techniques require control over both VMs, and are usually used in
experimentation rather than in practical attacks. We later refer to such
approaches as \emph{cooperative co-residency detection}.

\paragraph{Placement study in PaaS.}
While we mainly studied placement vulnerabilities in the context of
IaaS, we also experimented with Platform-as-a-Service (PaaS)
clouds. PaaS clouds offer elastic application hosting services.
Unlike IaaS where users are granted full control of a VM, PaaS
provides managed compute tasks (or instances) 
for the execution of hosted web applications, and allow multiple such instances
to share the same operating system. These clouds use either process-level
isolation via file system access controls, or increasingly Linux-style
containers (see~\cite{zhang2014cross} for a more detailed description). 
As such, logical side-channels alone are
usually sufficient for co-residency detection purposes. For instance, in PaaS clouds, co-resident instances often share the same public IP address as
the host machine. This is because the host-to-instance network is
often configured using Network Address Translation (NAT) and each
instance is assigned a unique port under the host IP address for
incoming connections. 

We found that many such logical side-channel-based co-residency
detection approaches worked on PaaS clouds, even on those using
containers. Specifically, we used both system-level interrupt
statistics via \texttt{/proc/interrupts} and shared public IP
addresses of the instances to detect co-location in
Heroku~\cite{heroku-network}. Note that both these techniques either
require direct access to victim instances, or a software vulnerability to
access \texttt{procfs} or initiate reverse connections, respectively.

Our brief investigation of co-location attacks in Heroku~\cite{heroku} showed that
\naive\ strategies like scaling two PaaS web applications to 30 instances with a
time interval of 5 minutes between them, resulted in co-location in 6 out of 10
attempts. Moreover, since the co-location detection was simple and quick including
the time taken for application scaling, we were able to do these experiments
free of cost. This result reinforces prior findings on PaaS co-location
attacks~\cite{zhang2014cross} and confirms the existence of cheap launch strategies to achieve
co-location and easy detection mechanisms to verify it. We do not investigate
PaaS clouds further in the rest of this paper.


\section{Threat Model}
\label{sec:threat-model}

Co-residency attacks in public clouds, as mentioned earlier, involve two steps:
a launch strategy and co-residency detection. We assume that the adversary has
access to tools to identify a set of target victims, and either knows victim VMs' launch
characteristics or can directly trigger their launches. The latter is possible by 
increasing load in order to cause the victim to scale up by launching more instances. The focus of
this study is to identify if there exists any launch strategy that an
adversary can devise to increase the chance of co-residency with a set of
targeted victim VMs.


In our threat model, we assume that the cloud provider is trusted and
the attacker has no affiliation of any form with the cloud
provider. This also means that the adversary has no internal knowledge
of the placement policies that are responsible for the VM placements in
the public cloud. An adversary also has the same interface for
launching and terminated VMs as regular customers, and no other
special interfaces. Even though there may be per-account limits on the
number of VMs that a cloud provider imposes, an adversary has access
to an unlimited number of accounts and hence has no limit on the
number of VMs he could launch at any given time. 

No resource-limited cloud provider is a match to an adversary with limitless
resources and hence realistically we assume that the adversary is
resource-limited. For the same reason, a cloud provider is vulnerable
to a launch strategy only when it is trivial or cost-effective for
an adversary. As such, we aim to (i) quantify the cost of executing a launch
strategy by an adversary, (ii) define a reference placement policy with which
the placement policies of real clouds can be compared, and (iii) define
metrics to quantify a placement vulnerability as existing when there are
cost-effective launch strategies that do better than they would against the
reference policy.

\paragraph{Cost of a launch strategy.} Quantifying the cost of a
launch strategy is straightforward: it is the cost of launching a number of VMs and
running tests to detect co-residency with one or more target
victim VMs. To be precise, the cost of a launch strategy $S$ is given
by $C_{S} = a \cdot P(a_{type}) \cdot T_{d}(v,a)$. Here $a$ is the number of
attacker VMs of type $a_{type}$ launched to get co-located with one of
the $v$ victim VMs. $P(a_{type})$ is the price of running one VM of
type $a_{type}$ for a unit time. $T_{d}(a,v)$ is the time (in billing units) to detect
co-residency between all pairs of $a$ attackers and $v$ victim VMs, excluding pairs
within each group. 
For simplicity, we assume that the attacker is
running all instances until the last co-residency check completes or that,
equivalently. When the time to finish co-residency checks is within the
granularity of one unit of billing time (e.g., one hour on EC2), this is
equivalent to a more refined model. 


\paragraph{Reference placement policy.}
In order to define placement vulnerability, we need a yardstick to
compare various placement policies and the launch strategies that
they may be vulnerable to.  To aid this purpose, we define a simple
reference placement policy that has good security properties against
co-residency attacks and use it to gauge the placement policies used
in public clouds. Let there be $N$ machines in a data center and let
each machine have unlimited capacity. Given a set of unordered VM
launch requests, the mapping of each VM to a machine follows a uniform
random distribution. Let
there be $v$ victim VMs assigned to $v$ unique machines among $N$,
where $v \ll N$. The probability of {\em at least one} collision
(i.e. co-residency) under the random placement policy and the above
attack scenario when attacker launches $a$ instances is given by $1 -
\big(1-v/N\big)^a$. We call this probability the reference
probability.\footnote{This probability event follows a hypergeometric
distribution.}
Recall that for calculating the cost of a
launch strategy under this reference policy, we also need to define
the price function, $P(vm_{type})$. For simplicity, we use the most
competitive minimum price offered by any cloud provider as the price
for the compute resource under the reference policy. 
For example, at the time of
this study, Amazon EC2 offered t2.small instances at \$0.026 per hour of instance
activity, which was the cheapest price across all three clouds considered in
this study.

\begin{figure*}[t]
\center
    \subfigure[GCE]{
      \includegraphics[scale=0.54]{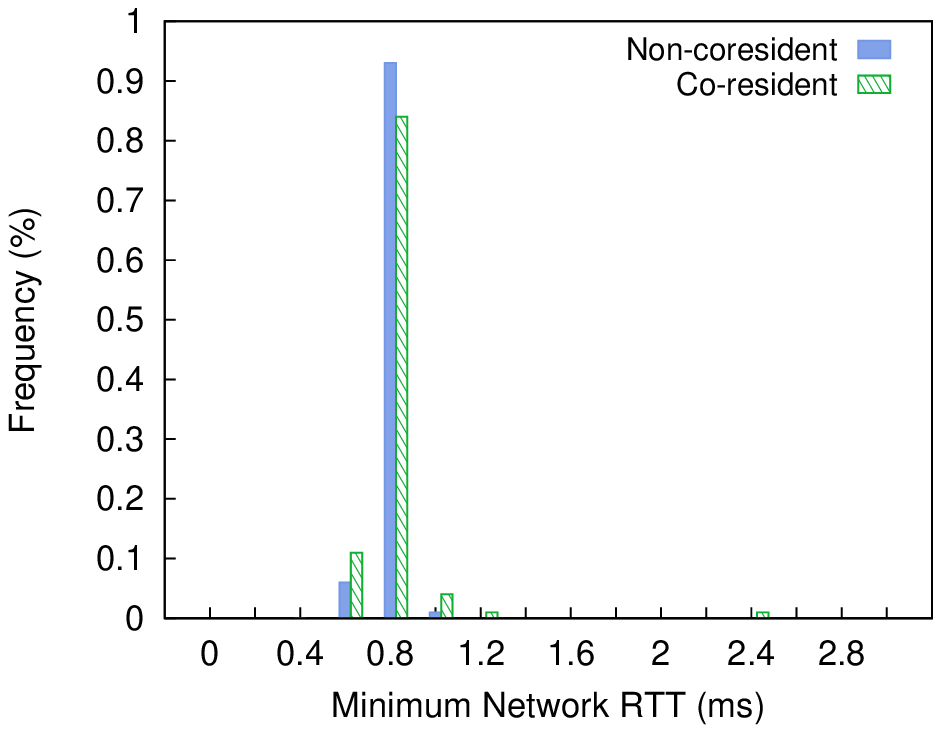}
    }
    \subfigure[EC2]{
      \includegraphics[scale=0.54]{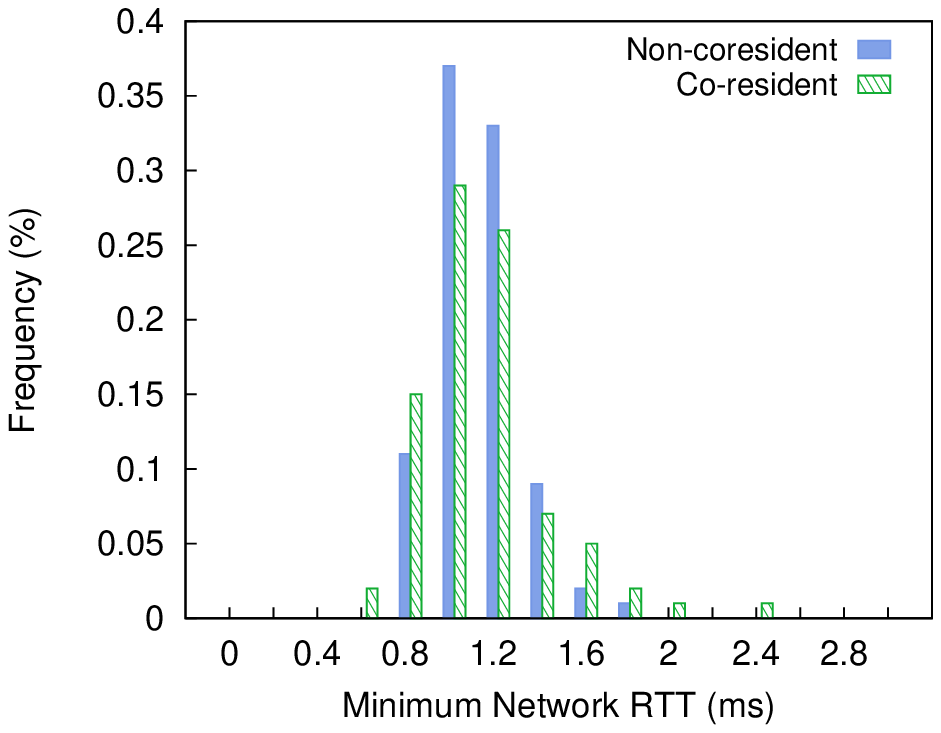}
    }
    \subfigure[Azure]{
      \includegraphics[scale=0.54]{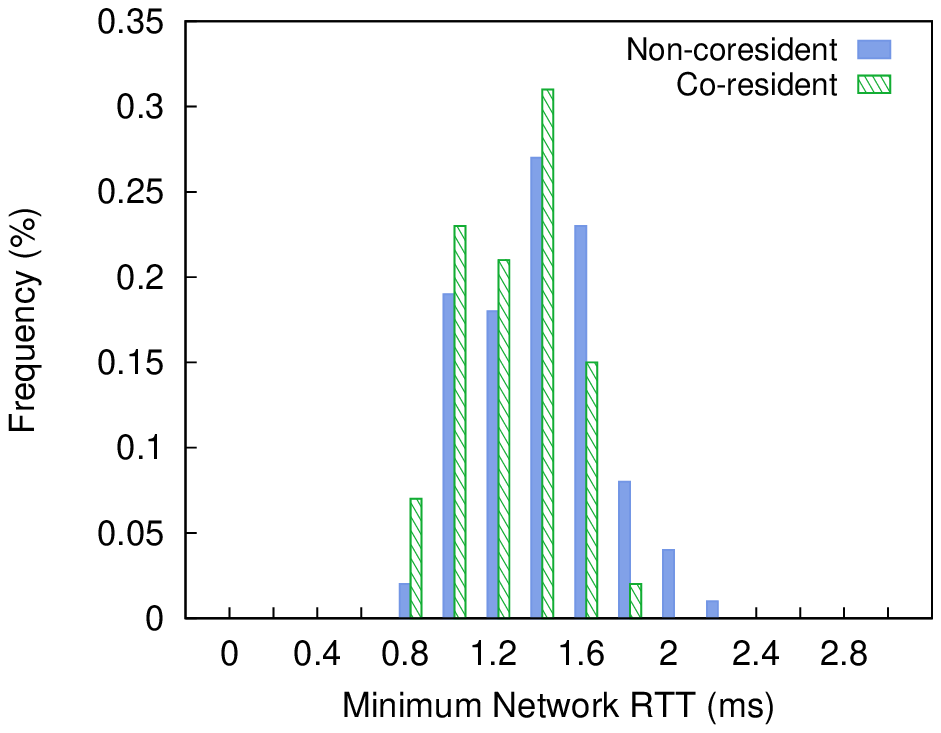}
    }
    \caption{{\bf Histogram of minimum network round trip times
        between  pairs of VMs.} The frequency is represented as a fraction
      of total number of pairs in each category. The figure does not
      show the tail of the histogram.}
    \label{fig:pings}
\end{figure*}

Note that the reference placement policy makes several simplifying assumptions,
but these only benefit the attacker. This is conservative as we will compare
our experimental results to the best possible launch strategy under the
reference policy. For instance, the assumption on unlimited
capacity of the servers only benefits the attacker as it never limits
the number of victim VMs an attacker could potentially co-locate
with. We use a conservative value of 1000 for $N$, which is
at least an order-of-magnitude less than the number of servers
(50,000) in the smallest reported Amazon EC2 data
centers~\cite{aws-reinvent-2014}. Similarly, the price function of
this placement policy also favors an attacker as it provides the
cheapest price possible in the market even though in reality a secure placement
policy may demand a higher price. Hence, it would be troubling if
the state-of-the-art placement policies used in public clouds does not
measure well even against such a conservative reference placement
policy.



\paragraph{Placement Vulnerability.} Putting it all together, we
define two metrics to gauge any launch strategy against a placement policy: (i) {\em normalized success
  rate}, and (ii) {\em cost-benefit}. The normalized success rate is the success
rate of the launch strategy in the cloud under test normalized to the success
rate of the same strategy under the reference placement policy. The cost-benefit
of a strategy is the additional cost that is incurred by the adversary in the
reference placement policy to achieve the same success rate as the strategy in the
placement policy under test. We define that a placement policy has a 
{\em placement vulnerability} if and only if there exists a launch strategy 
with a normalized success rate that is greater than 1.

Note that the normalized success rate quantifies how easy it is to get
co-location. On the other hand, the cost benefit metric helps to quantify how
cheap it is to get co-location compared to a more secure placement policy. These
metrics can be used to compare launch strategies under different placement
policies, where a higher value for any of these metrics indicate that the
placement policy is relatively more vulnerable to that launch strategy. An ideal
placement policy should aim to reduce both the success rate and the cost benefit
of any strategy.


\ignore{ 
\begin{figure}
\centering\
\footnotesize
  \begin{tabular}{|c|c|c|c|c|c|}
    \multicolumn{6}{c}{\bf GCE} \\
    \hline
     & \textbf{Mean} & \textbf{S.D.} & \textbf{Min} & \textbf{Max} & \textbf{Count}\\
    \hline
    NC & 0.69 & 0.17 & 0.38 & 17.62 & 85361 \\
    C & 0.68 & 0.14 & 0.54 & 2.22 & 179 \\
    \hline
    \multicolumn{6}{c}{\bf EC2} \\
    \hline
     & \textbf{Mean} & \textbf{S.D.} & \textbf{Min} & \textbf{Max} & \textbf{Count}\\
    \hline
    NC & 1.57 & 2.47 & 0.42 & 30.01 & 57169 \\
    C & 2.34 & 3.49 & 0.55 & 17.17 & 149 \\
    \hline
    \multicolumn{6}{c}{\bf Azure} \\
    \hline
     & \textbf{Mean} & \textbf{S.D.} & \textbf{Min} & \textbf{Max} & \textbf{Count}\\
    \hline
    NC & 0.49 & 4.64 & 1.29 & 0.3 & 45878 \\
    C & 0.58 & 2.34 & 1.17 & 0.26 & 1099 \\
    \hline
  \end{tabular}
  \caption{\textbf{Distribution of Minimum Network Round Trip Times Across All Pairs.}
    All latencies are in milliseconds. Count is the total number of
    pairs in each category, and NC and C denote non-coresident and
    co-resident pairs, respectively. }
\label{tab:pings}
\end{figure}
} 

\section{Detecting Co-Residence}
\label{sec:cores-test}

An essential prerequisite for the placement vulnerability study is access to a
co-residency detection technique that identifies whether two VMs are resident on
the same physical machine in a third-party public cloud.

\paragraph{Challenges in modern clouds.}
Applying the detection techniques mentioned in~\secref{sec:background} is no
longer feasible in modern clouds. In part due to the vulnerability disclosure by
Ristenpart et al.~\cite{rist:hey-you:ccs:2009}, modern public clouds have
adopted new technologies that enhance the isolation between cloud tenants and
thwart known co-residence detection techniques. In the network layer, virtual
private clouds (VPCs) have been broadly employed for data center
management~\cite{beach2014vpc,aws-vpc-whitepaper}. With VPCs, internal IP
addresses are private to a cloud tenant, and can no longer be used for cloud
cartography. Although EC2 allowed this in older generation instances (called
EC2-classic), this is no longer possible under Amazon VPC setting. In addition,
VPCs require communication between tenants to use public IP addresses for
communication. As shown in \figref{fig:pings}, the network timing test is also
defeated, as using public IP addresses seems to involve routing in the data
center network rather than short-circuiting through the hypervisor. Here, the
ground-truth of co-residency is detected using memory-based covert-channel
(described later in this section). Notice that there is {\em no} clear
distinction between the frequency distribution of the network round trip times
of co-resident and non-coresident pairs on all three clouds.

In the system layer, persistent storage using local disks is no longer
the default.  For instance, many Amazon EC2 instance types do not
support local storage~\cite{ec2-instance-store}; GCE and Azure provide only
local Solid State Drives (SSD)~\cite{gce-disks,azure-disks}, which are
less susceptible to detectable delays from long seeks. In addition,
covert channels based on last-level
caches~\cite{rist:hey-you:ccs:2009,venkat:rfa:ccs2012,xu2011exploration,
  zhang11:homealone} are less reliable in modern clouds that use
multiple CPU packages.  Two VMs  sharing the same machine may
not share LLCs to establish the covert channel. Hence, these LLC-based
covert-channels can only capture a subset of co-resident
instances.

As a result of these technology changes, none of the prior techniques
for detecting co-residency reliably work in modern clouds, compelling
us to develop new approaches for our study.

\subsection{Co-residency Tests}
\label{ssec:corestest}



We describe in this subsection a pair of tools for co-residency tests,
with the following design goals:


\newenvironment{packeditemize}{
\begin{list}{$\bullet$}{
\setlength{\labelwidth}{8pt}
\setlength{\itemsep}{0pt}
\setlength{\leftmargin}{\labelwidth}
\addtolength{\leftmargin}{\labelsep}
\setlength{\parindent}{0pt}
\setlength{\listparindent}{\parindent}
\setlength{\parsep}{0pt}
\setlength{\topsep}{3pt}}}{\end{list}}

\begin{packeditemize}

\item Applicable to a variety of {\em heterogeneous} software and
hardware stacks used in public clouds.


\item Detect co-residency with {\em high confidence}: the false
  detection rate should be low even in the presence of background
  noise from other neighboring VMs.

\item Detect co-residency {\em fast} enough to
facilitate experimentation among large sets of VMs.

\end{packeditemize}

We chose a performance covert-channel based detection technique that exploits shared
hardware resources, as this type of covert-channels are often hard to remove and
most clouds are very likely to be vulnerable to it.

A covert-channel consists of a {\em sender} and a {\em receiver}. The sender
creates contention for a shared resource and uses it to signal another
tenant that potentially share the same resource. The receiver, on the other hand, senses
this contention by periodically measuring the performance of that shared
resource. A significant performance degradation measured at the receiver results
in a successful detection of a sender's signal. Here the reliability of the
covert-channel is highly dependent on the choice of the shared resource and the
level of contention created by the sender. The sender is the key component of
the co-residency detection techniques we developed as part of this study. 

\begin{figure}[h]
\center
\scriptsize
\begin{verbbox}
// allocate memory multiples of 64 bits
char_ptr = allocate_memory((N+1)*8) 
//move half word up
unaligned_addr = char_ptr + 2 

loop forever:
   loop i from (1..N):
     atomic_op(unaligned_addr + i, some_value)
   end loop
end loop
\end{verbbox}
\theverbbox
\caption{
  \label{fig:mlock-code} {\bf Memory-locking -- Sender.}}
\vspace{-0.1in}
\end{figure}

\paragraph{Memory-locking sender.} Modern x86 processors support atomic
 memory operations, such as \texttt{XADD} for atomic addition, and maintain their
 atomicity using cache coherence protocols. However, when a locked operation
 extends across a cache-line boundary, the processor may lock the memory bus
 temporarily~\cite{wu2012whispers}. This locking of the bus can be detected as
 it slows down other uses of the bus, such as fetching data from DRAM. Hence,
 when used properly, it provides a timing covert channel to send a signal to another
 VM. Unlike cache-based covert channels, this technique works regardless of
 whether VMs share a CPU core or package.

We developed a sender exploiting this shared memory-bus covert-channel. The
psuedocode for the sender is shown in~\figref{fig:mlock-code}. The sender
creates a memory buffer and uses pointer arithmetic to force atomic operations
on unaligned memory addresses. This indirectly locks the memory bus even on all
modern processor architectures~\cite{wu2012whispers}.

\begin{figure}[h]
\centering
\scriptsize
\begin{verbbox}
size = LLC_size * (LLC_ways +1)
stride = LLC_sets * cacheline_sz)
buffer = alloc_ptr_chasing_buff(size, stride)
                                
loop sample from (1..10): //number of samples
     start_rdtsc = rdtsc()
     loop probes from (1..10000):
         probe(buffer); //always hits memory
     end loop
     time_taken[sample] = (rdtsc() - start_rdtsc)
end loop
\end{verbbox}
\theverbbox
\caption{
  \label{fig:mprobe-code} {\bf Memory-probing -- Receiver.}}
\vspace{-0.1in}
\end{figure}

\paragraph{Receivers.} With the aforementioned memory-locking sender, there are
several ways to sense the memory-locking contention induced by the sender in
another co-resident tenant instance. All the receivers measure the
memory bandwidth of the shared system. We
present two types of receivers that we used in this study that works on
heterogeneous hardware configurations. 

\vspace{.6em}\noindent\textit{Memory-probing} receiver uses carefully crafted memory requests
that always miss in the cache hierarchy and always hit memory. 
This is ensured by constricting the data accesses of the receiver
into a {\em single LLC set}. 
In order to
evade hardware prefetching, we use a pointer-chasing buffer to randomly access a
list of memory addresses (pseudocode shown in~\figref{fig:mprobe-code}). The
time needed to complete a fixed number of probes (e.g., 10,000) provides a
signal of co-residence: when the sender is performing locked operations, loads
from memory proceed slowly.

\vspace{.6em}\noindent\textit{Memory-locking} receiver is similar to the sender but measures the
number of unaligned atomic operations that could be completed per unit
time. Although it also measures the memory bandwidth, unlike the memory-probing
receiver, it works even when the cache architecture of the machine is unknown.\\

The sender along with these two receivers form our two novel co-residency
detection methods that we use in this study: {\em memory-probing test} and {\em
  memory-locking test} (named after their respective receivers). These comprise
our co-residency test suite. Each test in the suite starts by running the
receiver on one VM while keeping the other idle. The performance measured by
this run is the baseline performance without contention. Then the receiver and
the sender are run together. If the receiver detects decreased performance, the
tests conclude that the two VMs are co-resident. We use a slowdown threshold to
detect when the change in receiver performance indicates co-residence (discussed
later in the section).

\ignore{ 
\begin{figure}
\centering\
\footnotesize
\begin{tabular}{|c|c|c|c|c|}
  \hline
  \textbf{Machine} & \textbf{Clock} & \textbf{SMT} & \textbf{LLC} & \textbf{Memory}\\ 
  \textbf{Architecture} & (GHz) & \textbf{Cores} & (Ways x Set) & \textbf{Architecture} \\
  \hline  
  Core i5-4570 & 3.20 & 4 & 12 x 8192 & UMA \\
  Core i7-2600 & 3.40 & 8 & 18 x 8192 & UMA \\
  Xeon E5645 & 2.40 & 6 & 16 x 12288 & UMA \\
  Xeon X5650 & 2.67 & 12 & 16 x 12288 & NUMA \\
  \hline
\end{tabular}
\caption{\textbf{Local Testbed Machine Configuration.} All are Intel
  machines. SMT stands for Simultaneous Multi-Threaded cores (in
  Intel parlance, Hyper-threads). Ways x Sets x Word Size gives the
  LLC size. The word size is 64 bytes on all these x86-64 machines.}
\label{tb:mach-config}

\end{figure}
} 

\begin{figure}[h]
\vspace{-0.12in}
\centering
\footnotesize
\begin{tabular}{|c|c|c|c|c|c|}
  \hline
  \textbf{Machine} & \textbf{Cores} & \textbf{Memory} & \textbf{Memory} & \textbf{Socket}\\ 
  \textbf{Architecture} & & \textbf{Probing} & \textbf{Locking} & \\
  \hline  
  Xeon E5645 & 6 & 3.51 & 1.79 & Same\\
  \hline
  Xeon X5650 & 12 & 3.61 & 1.77 & Same\\
  Xeon X5650 & 12 & 3.46 & 1.55 & Diff.\\
  \hline
\end{tabular}
\caption{\textbf{Memory-probing and -locking on testbed machines.}
  Slowdown relative to the baseline performance observed
  by the receiver averaged across 10 samples. Same -- sender and
  receiver on different cores on the same socket, Diff. -- sender and
  receiver on different cores on different sockets. Xeon E5645 machine
  had a single socket.}
\vspace{-0.15in}
\label{tb:local-perf}
\end{figure}

\paragraph{Evaluation on local testbed.} In order to measure the efficacy
 of this covert-channel we ran tests in our local testbed. Results of
 running memory-probing and -locking tests under various configurations are shown
 in~\figref{tb:local-perf}. The hardware architectures of these machines are
 similar to what is observed in the cloud~\cite{money-socc12}. Across these
 hardware configurations, we observed a performance degradation of at least 3.4$\times$
 compared to not running memory-locking sender on a non-coresident instance
 (i.e. a baseline run with idle sender), indicating reliability. Note that this
 works even when the co-resident instances are running on cores on different
 sockets, which does not share the same LLC (works on {\em heterogeneous}
 hardware). Further, a single run takes one tenth of a second to complete and
 hence is also {\em quick}. 

Note that for this test suite to work in the real world, an attacker requires
control over both the VMs under test, which includes the victim. We call this
scenario as co-residency detection under cooperative victims (in short, {\em
  cooperative co-residency detection}).  Such a mechanism is
sufficient to observe placement behavior in public clouds (\secref{ssec:coop-test}). We further investigated approaches
to detect co-residency under a realistic setting with an uncooperative
victim. In~\secref{ssec:uncoop-test} we show how to adapt
 the memory-probing test  to detect co-location with one of the many
webservers behind a load balancer.

\vspace{-0.05in}
\subsection{Cooperative Co-residency Detection}
\label{ssec:coop-test}
\vspace{-0.05in}

In this section, we describe the methodology we used to detect co-residency in
public clouds. For the purposes of studying placement policies, we had the
flexibility to control both VMs that we test for co-residence. We did this by
launching VMs from two separate accounts and test them for pairwise
co-residence. We encountered several challenges when running the co-residency
test suite on three
different public clouds - Google Computer Engine, Amazon EC2 and Microsoft
Azure.

First, we had to handle noise from neighboring VMs sharing the same
host. Second, hardware and software heterogeneity in the three different public
clouds required special tuning process for the co-residency detection
tests. Finally, testing co-residency for a large set of VMs demanded a scalable
implementation. We elaborate on our solution to these challenges below.

\paragraph{Handling noise.} 
Any noise from neighboring VMs could affect the performance of the receiver with
and without the signal (or baseline) and result in misdetection. To handle such
noise, we alternate between measuring the performance with and without the
sender's signal, such that any noise equally affects both the
measurements. Secondly, we take ten samples of each measurement and only detect
co-residence if the ratios of {\em both} the mean and median of these samples
exceed the threshold. As each run takes a fraction of a second to complete,
repeating 10 times is still fast enough.



\begin{figure}[t]
\vspace{-0.12in}
\center
\footnotesize
\begin{tabular}{|c|c|c|c|c|}
  \hline
  \textbf{Cloud} & \textbf{Machine} & \textbf{Clock} & \textbf{LLC} \\ 
  \textbf{Provider} & \textbf{Architecture} & (GHz) & (Ways $\times$ Set) \\
  \hline  
  EC2 & Intel Xeon E5-2670 & 2.50 & 20 $\times$ 20480 \\
  GCE & Generic Xeon* & 2.60* & 20 $\times$ 16384 \\
  Azure & Intel E5-2660 & 2.20  & 20 $\times$ 16384 \\
  Azure & AMD Opteron 4171 HE & 2.10 & 48 $\times$ 1706 \\
  \hline
\end{tabular}
\caption{\textbf{Machine configuration in public clouds.} The machine
  configurations observed over all runs with small instance types. GCE
  did not reveal the exact microarchitecture of the physical
  host~(*). Ways $\times$ Sets $\times$ Word Size gives the LLC size. The word size
  for all these x86-64 machines is 64 bytes.}
\label{tb:cloud-mach-config}
\vspace{-0.12in}
\end{figure}

\begin{figure*}[t]
\center
    \subfigure[GCE]{
      \includegraphics[scale=0.54]{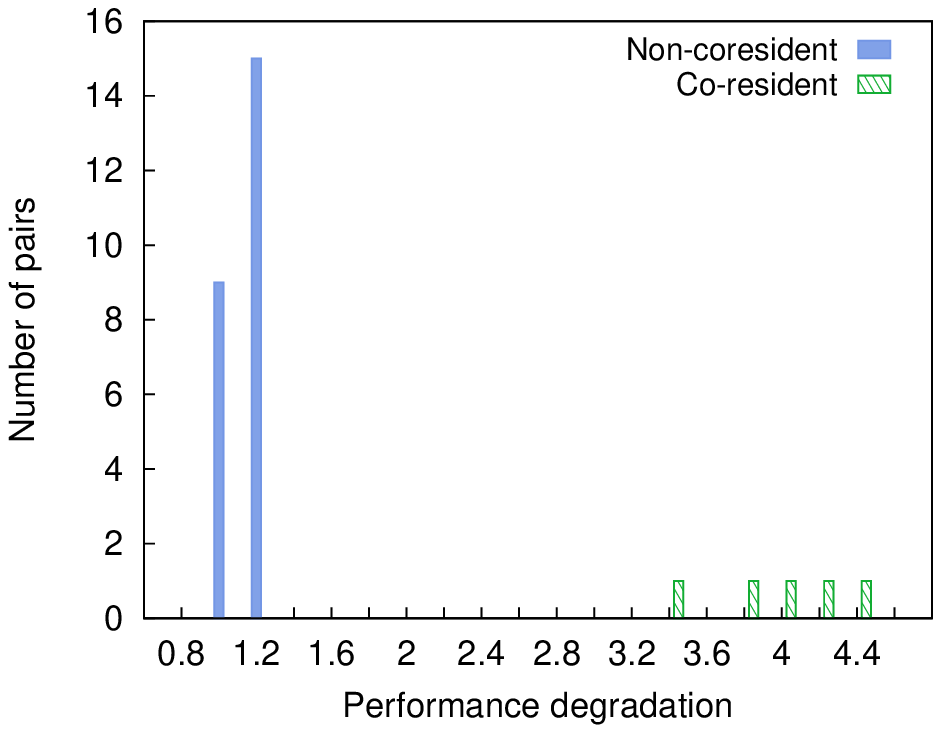}
    }
    \subfigure[EC2]{
      \includegraphics[scale=0.54]{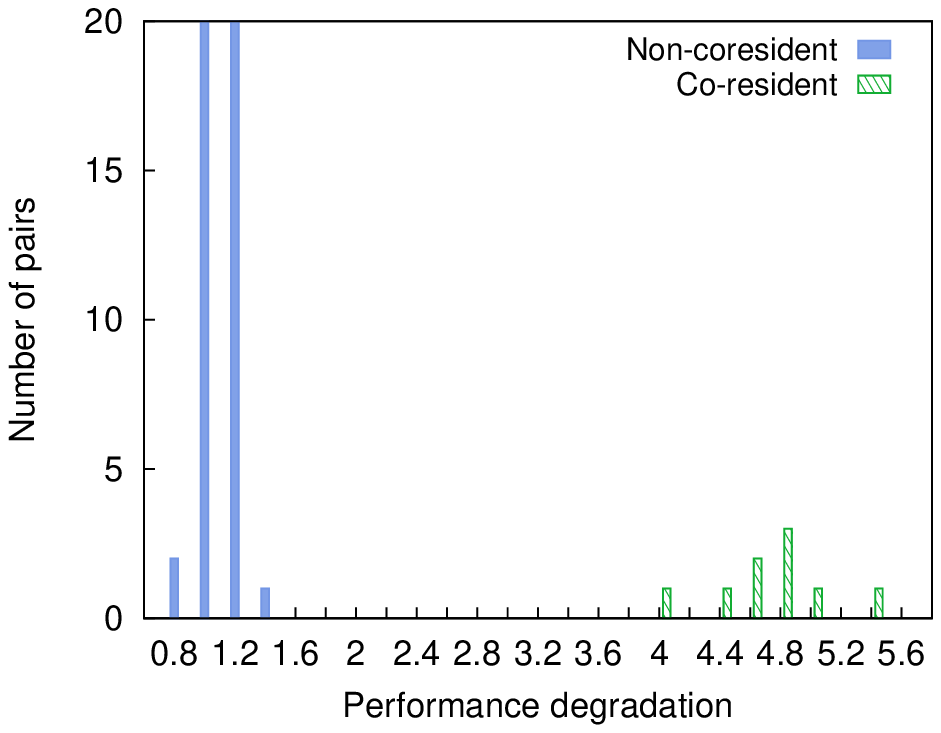}
    }
    \subfigure[Azure -- Intel machines]{
      \includegraphics[scale=0.54]{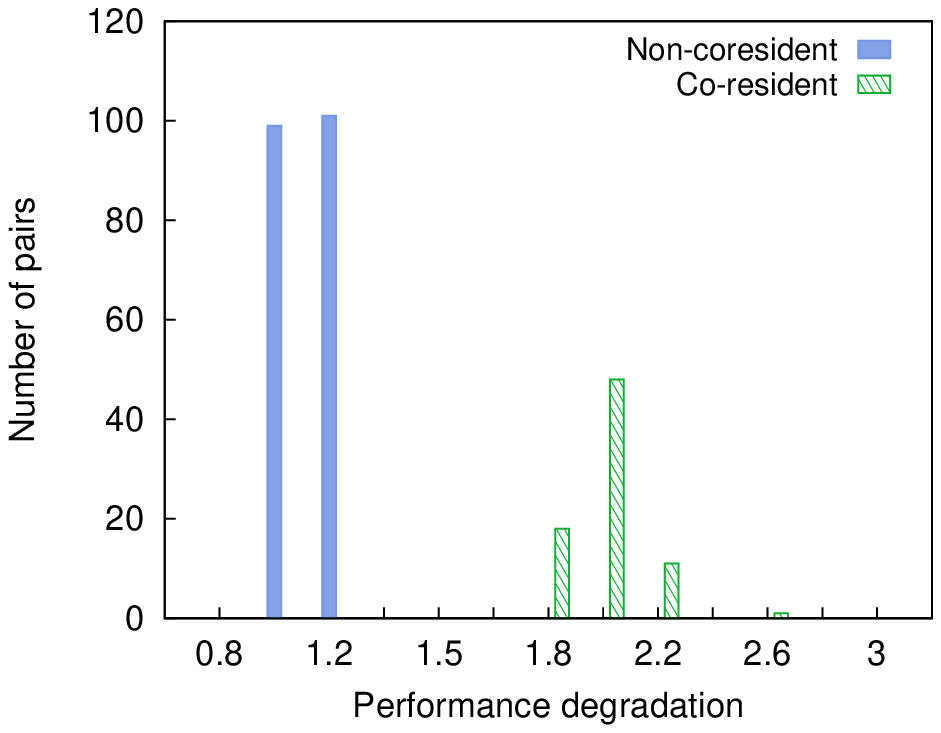}
    }
    \vspace{-0.15in}
    \caption{\textbf{Distribution of performance degradation of
        memory-probing test.}  For
      varying number of pairs on each cloud (GCE:29, EC2:300,
      Azure:278). Note the x-axis plots performance degradation. Also for EC2
      x-axis range is cut short at 20 pairs for clarity.}
    \label{fig:mprobe-scores}
\end{figure*}

\paragraph{Tuning thresholds.}
As expected, we encountered different machine configurations on the three
different public clouds (shown in~\figref{tb:cloud-mach-config}) with
heterogeneous cache dimensions, organizations and replacement
policies~\cite{ivy-bridge-llc,sandy-bridge-llc}. This affects the performance
degradation observed by the receivers with respect to the baseline and the ideal
threshold for detecting co-residency. This is important because the thresholds we use to detect co-residence yield
false positives, if set too low, and false negatives if set too high. Hence, we
tuned the threshold to each hardware we found on all three clouds.



We started with a conservative threshold of 1.5$\times$ and tuned
to a final threshold of 2$\times$ for GCE and EC2 and 1.5$\times$ for Azure for both the
memory-probing and -locking tests. \figref{fig:mprobe-scores} shows the distribution of performance
degradation under the memory-probing tests across Intel machines in EC2, GCE,
and Azure. For GCE and EC2, a performance degradation threshold of 2 clearly
separates co-resident from non-coresident instances. For all Intel machines we
encountered, although we ran both memory-locking and -probing tests, 
memory-probing was sufficient to detect co-residency. For Azure, overall we observe
lower performance degradation and the initial threshold of 1.5 was sufficient to
detect co-location on Intel machines.

\begin{figure}[h]
  \center
  \vspace{-0.1in}
  \includegraphics[scale=0.75]{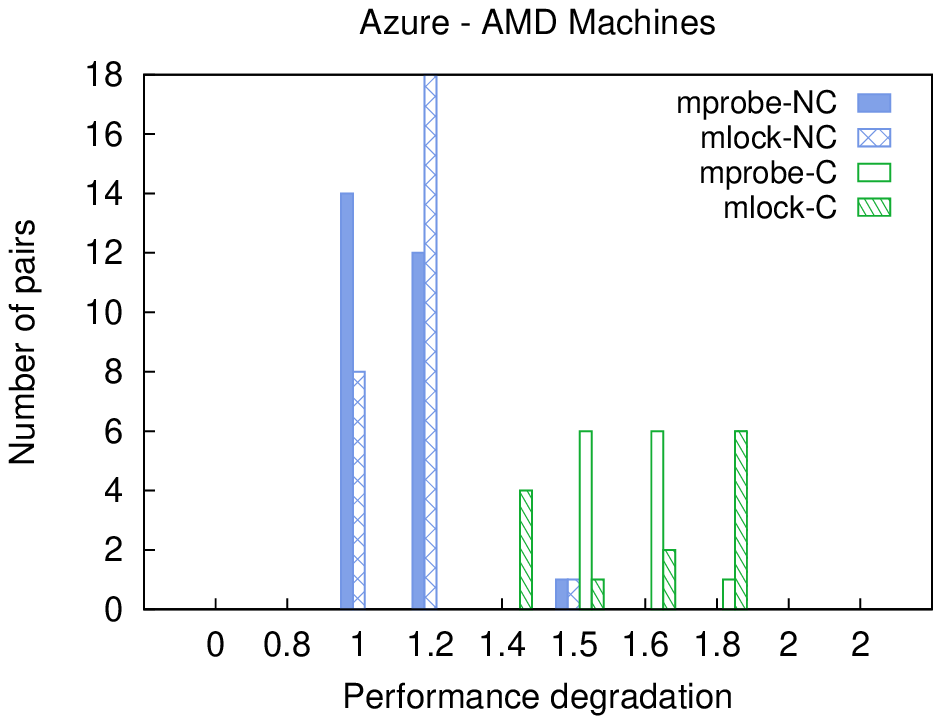}
  \caption{{\bf Distribution of performance degradation of memory-probing and
      -locking tests.} On AMD machines in Azure with 40 pairs of
    nodes. Here NC stands for non-coresident and C, co-resident pairs. Note that
     the x-axis plots performance degradation.}
  \label{fig:azure-amd}
\end{figure}

The picture for AMD machines in Azure differs significantly as shown
in \figref{fig:azure-amd}. The distribution of performance degradation
for both memory-locking and memory-probing shows that, unlike for
other architectures, co-residency detection is highly sensitive to the
choice of the threshold for AMD machines. This may be due to the more
associative cache (48 ways vs. 20 for Intel), or different handling of
locked instructions. For these machines, a threshold of 1.5 was high
enough to have no false positives, which we verified by hand checking the
instances using the two covert-channels and observed consistent performance
degradation of at least 50\%. We determine a pair of VMs as
co-resident if the degradation in {\em either} of the tests is above
this threshold. 
We did not detect any cross-architecture (false) co-residency
detection in any of the runs.

\paragraph{Scaling co-residency detection tests.}
Testing co-residency at scale is time-consuming and increases quadratically with
the number of instances: checking 40 VM instances, involves 780 pair-wise
tests. Even if each run of the entire co-residency test suite takes only 10
seconds, a \naive\ sequential execution of the tests on all the pairs will take
2 hours. Parallel co-residency checks can speed checking, but concurrent tests
may interfere with each other.


To parallelize the test, we partition the set of all VM pairs ($\choose{v+a}{2}$)
into sets of pairs with no VMs twice; we run one of these sets at a time and
record which pairs detected possible co-residence. After running all sets, we
have a set of candidate co-resident pairs, which we test
sequentially. Parallelizing co-residency tests significantly decreased the time
taken to test all co-residency pairs. For instance, the parallel version of the
test on one of the cloud providers took 2.4 seconds per pair whereas the serial
version took almost 46.3 seconds per pair (a speedup of 20x). While there are
faster ways to parallelize co-residency detection, we chose this approach for simplicity.

\paragraph{Veracity of our tests.}
Notice that a performance degradation of 1.5x, 2x and 4x corresponds to 50\%,
100\% and 300\% performance degradation. Such high performance degradation
(even 50\%) is clear enough signal to declare co-residency due to resource
sharing. Furthermore, we hand checked by running the two tests in isolation on the detected instance-pairs for a
significant fraction of the runs for all clouds and
observed a consistent covert-channel signal. Thus our methodology
did not detect any false positives, which are more detrimental to our study than
false negatives. Although {\em co-residency} here implies sharing of memory
channel, which may not always mean sharing of cores or other per-core hardware
resources.

\subsection{Co-residency Detection on Uncooperative Victims}
\label{ssec:uncoop-test}
Until now, we described a method to detect co-residency with a cooperative
victim. In this section, we look at a more realistic setting where an adversary
wishes to detect co-residency with a victim VM with accesses limited to only
public interfaces like HTTP or a key-value (KV) store's put-get interface. We show that the basic cooperative co-residency detection  can also be employed to detect co-residency with an
uncooperative victim in the wild.


\begin{figure}
\center
\includegraphics[scale=0.40]{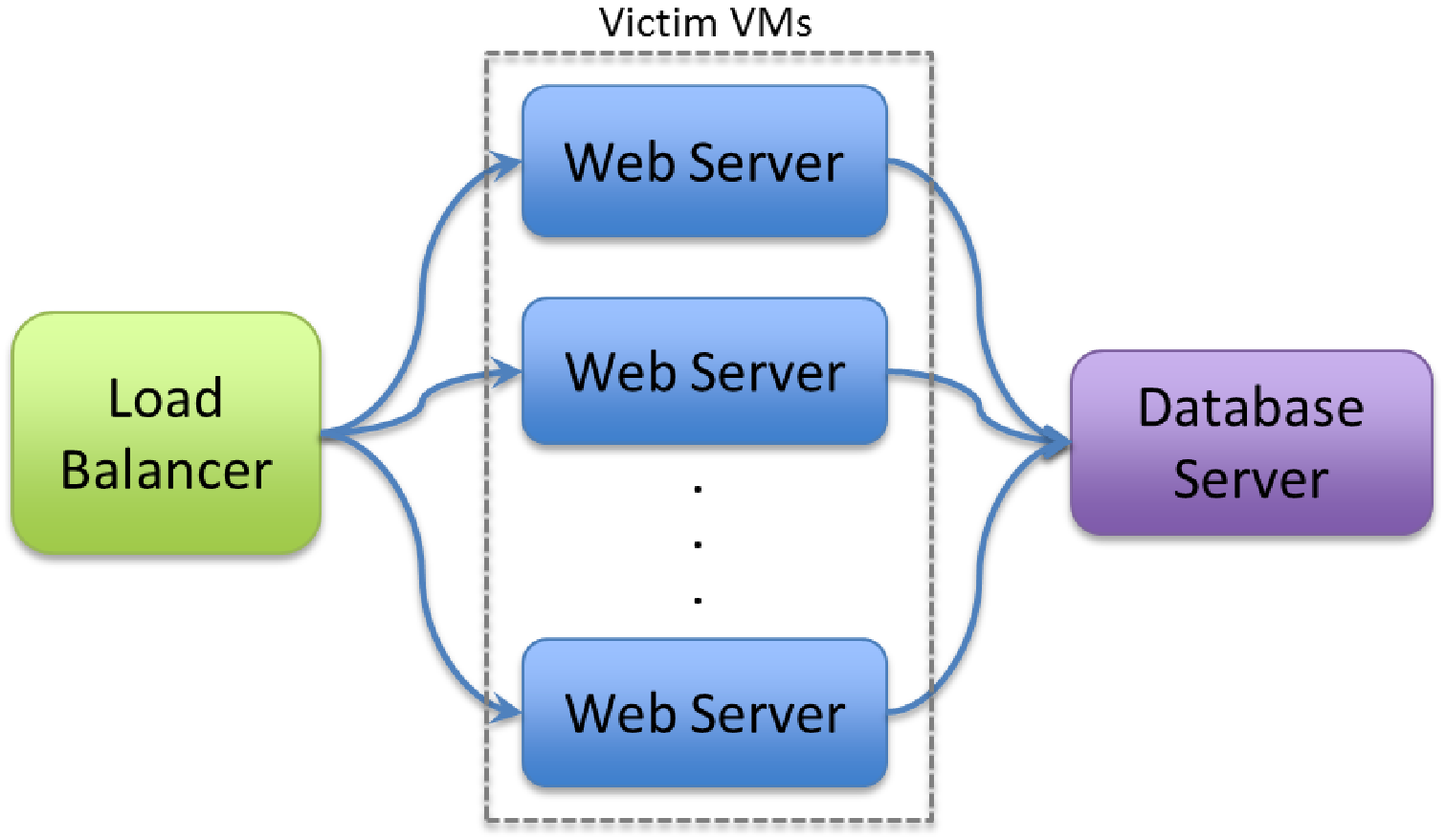}
\caption{{\bf An example victim web application architecture.}}
    \label{fig:web-app-arch}
\end{figure}

\paragraph{Attack setting.}
Unlike previous attack scenarios, we assume the attacker has no access to the
victim VMs or its application other than what is permitted to any user on the
Internet. That is, the victim application exposes a well-known public interface
(e.g., HTTP, FTP, KV-store protocol) that allows incoming requests, which is also
the only access point for the attacker to the victim. The front end of this
victim application can range from caching or data storage services (e.g.,
memcached, cassandra) to generic webservers. We also assume that there may be multiple instances of this
front-end service running behind a load balancer.  Under
this scenario, the attacker wishes to detect co-location with one or more of
the front-facing victim VMs.


\paragraph{Co-residency test.}
We adapt the memory tests used in previous section by running the memory-locking
sender in the attacker instance. For a receiver, we use the public interface
exposed by the victim by generating a set of requests that potentially makes the
victim VMs hit the memory bus. This can be achieved by looping through a large number
of requests of sizes approximately equal or greater than the size of the
LLC. This creates a performance side-channel that leaks co-residency
information. This receiver runs in an independent VM under the adversary's
control, which we call the co-residency detector.


\paragraph{Experiment setup.}
To evaluate the efficacy of this method, we
used the Olio multi-tier web application~\cite{olio-workload} that is
designed to mimic a social-networking application. We used an instance of this
workload from CloudSuite~\cite{cloudsuite2}. Although Olio
supports several tiers (e.g., memcached to cache results of database queries),
we configured it with two tiers as shown in~\figref{fig:web-app-arch}, with
each webserver and the database server running in a separate VM of type t2.small
on Amazon EC2. Multiple of these webserver VMs are configured behind a
HAProxy-based load balancer~\cite{haproxy} running in an m3.medium instance (for
better networking performance). The load balancer follows the standard
configuration of using round-robin load balancing algorithm with sticky client
sessions using cookies. We believe such a victim web application and its
configuration is a reasonable generalization of real world applications running
in the cloud.

For the attacker, we use an off-the-shelf HTTP performance measurement utility
called \texttt{HTTPerf}~\cite{httperf} as the receiver in the co-residency
detection test. This receiver is run inside a t2.micro instance (for free of
charge). We used a set of 212 requests that included web pages and web objects
(images, PDF files). We gathered these requests
using the access log of manual navigation around the web application from a web browser.

\paragraph{Evaluation methodology.}
We start with a known co-resident VM pair using the cooperative co-residency
detection method. We configure one of the VMs as a victim webserver VM and
launch four more VMs: two webservers, one database server and a load balancer,
all of which are not co-resident with the attacker VM. 


Co-residency detection starts by measuring the average request latency at the
receiver inside the co-residency detector for the baseline (with idle attacker)
and contended case with the attacker running the memory-locking sender. A
significant performance degradation between the baseline and the contended case
across multiple samples reveal co-residency of one of the victim VMs with the
attacker VM. On Amazon EC2, with the above setup we observed an average request
latency of $4.66$ms in the baseline case and a $10.6$ms in the memory-locked
case, i.e., a performance degradation of $\approx 2.3\times$.

\begin{figure}[t]
\center
\includegraphics[scale=0.75]{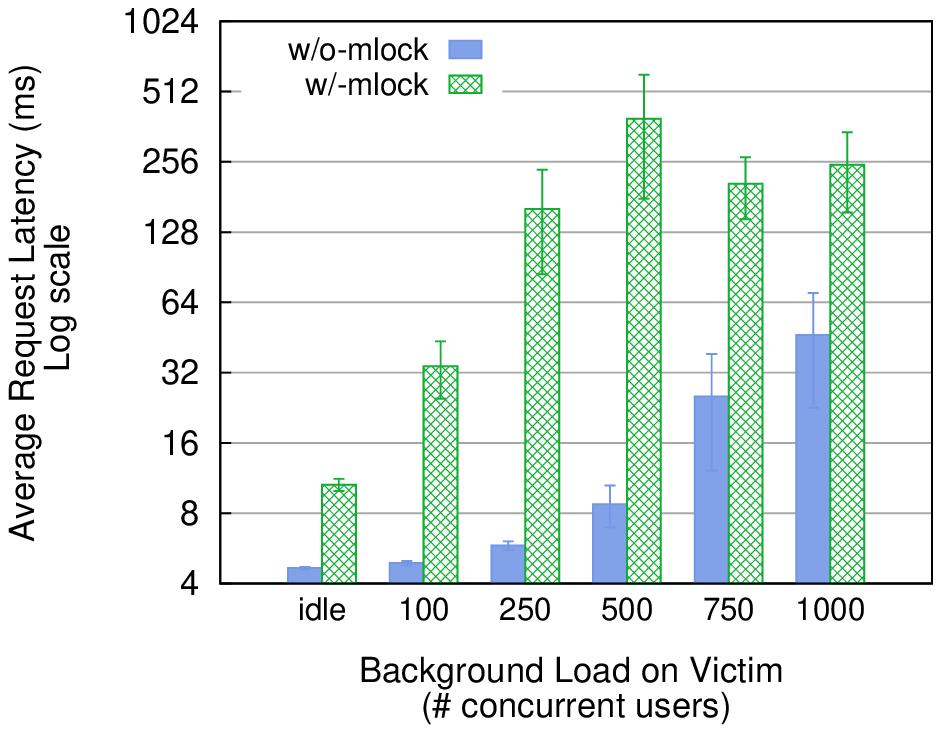}
\caption{{\bf Co-residency detection on an uncooperative victim.} The
  graph shows the average request latency at the co-residency detector
  without and with memory-locking sender running on the co-resident
  attacker VM under varying background load on the victim. Note that
  the y-axis is in log scale. The load is in the number of concurrent
  users, where each user on average generates 20 HTTP requests per
  second to the webserver.}
    \label{fig:uc-vary-load}
\end{figure}

\paragraph{Background noise.} The above test was performed when the victim web
application was idle. In reality, any victim in the cloud might experience constant or
varying background load on the system. False positives or negatives may occur
when there is spike in load on the victim servers. In such case, we use the same
solution as in~\secref{ssec:coop-test} --- alternating between measuring the idle
and the contended case.

In order to gauge the efficacy of the test under constant background load, we
repeated the above experiment with varying load on the victim. The result of
this experiment is summarized in~\figref{fig:uc-vary-load}. Counterintuitively,
we found that a constant load on the background server exacerbates the
performance degradation gap, hence resulting in a clearer signal of
co-residency. This is because running memory-locking on the co-resident attacker
increases the service time of all requests as majority of the requests rely on
memory bandwidth. This increases the queuing delay in the system and in turn
increasing the overall request latency. Interestingly, this aforementioned
performance gap stops widening at higher loads of 750 to 1000 concurrent users
as the system hits a bottleneck (in our case a network bottleneck at the load
balancer) even without running the memory-locking sender. Thus, detecting
co-residency with a victim VM that
is part of a highly loaded and bottlenecked application would be hard using this test.

\ignore{
\begin{figure}
\centering\
\footnotesize
  \begin{tabular}{|c|c|c|c|c|c|}
    \hline
    \textbf{Load}&\textbf{Thruput} & \multicolumn{3}{c|}{\textbf{Avg. CPU Util.} (\%)} & \textbf{N/w b/w} \\
    (users) & (ops/sec) & \textbf{WS} & \textbf{DB} & \textbf{LB} & (MB/s)\\
    \hline
    100 & 19.89 & 4 & 6 & 10 & 7.34\\
    250 & 49.99 & 10 & 13 & 24 & 18.48\\
    500 & 100.47 & 19 & 25 & 43 & 35.88\\
    750 & 151.78 & 27 & 38 & 52 & 53.58\\
    1000 & 181.42 & 31 & 45 & 54 & 64.18\\
\hline
\end{tabular}
\caption{}
\label{fig:olio-baseline}
\end{figure}
}

\begin{figure}[h]
\center
\includegraphics[scale=0.75]{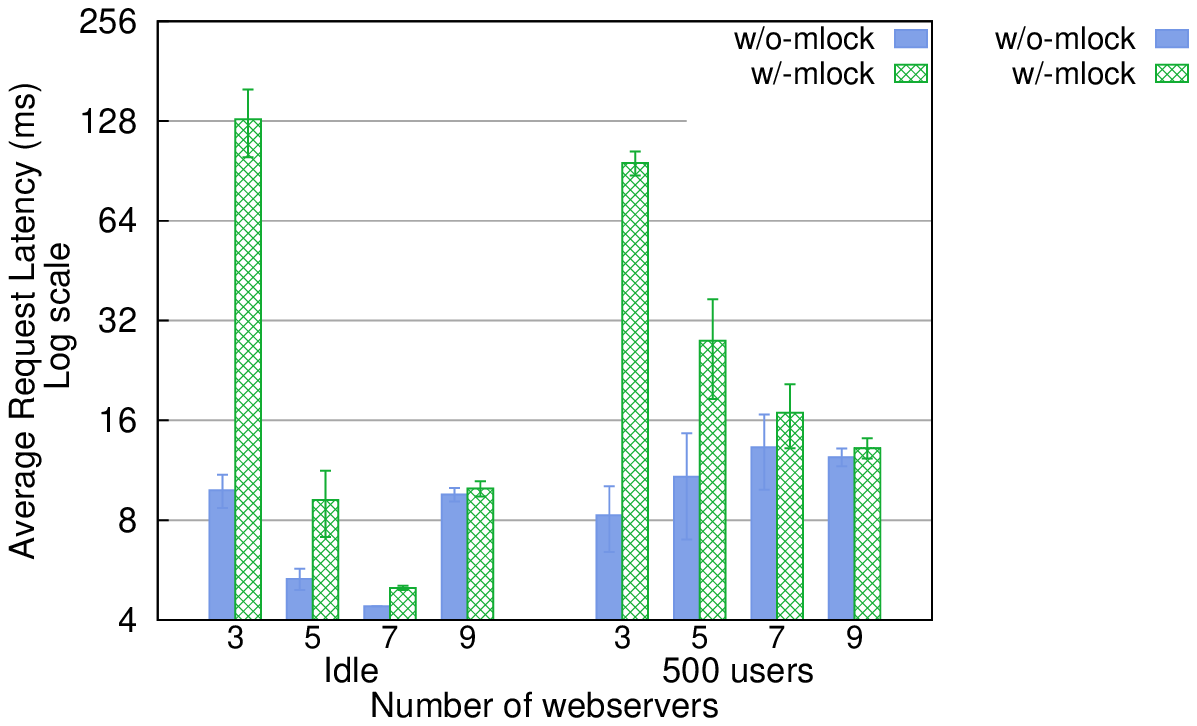}
\caption{{\bf Varying number of webservers behind the load balancer.}  The graph
  shows the average request latency at the co-residency detector without and
  with memory-locking sender running on the co-residency attacker VM under
  varying background load on the victim. Note that the y-axis is in log
  scale. The error bars show the standard deviation over 5 samples.}
    \label{fig:uc-vary-server-load}
\end{figure}

We also experimented with increasing the number of victim webservers behind the
load balancer beyond 3 (\figref{fig:uc-vary-server-load}). As expected, the
co-residency signal grew weaker with increasing victims, and at 9
webservers, the performance degradation was too low to be useful for
detecting co-residency. 



\section{Placement Vulnerability Study}
\label{sec:results}

In this section, we evaluate three public clouds, Amazon EC2, Google Compute Engine
and Microsoft Azure, for placement
vulnerabilities and answer the following questions:
(i)~what are all the strategies that an adversary can employ to
increase the chance of co-location with one or more victim VMs? (ii)~what are
the chances of success and cost of each strategy? and (iii)~how do these
strategies compare against the reference placement policy
introduced in~\secref{sec:threat-model}?





\subsection{Experimental Methodology}

Before presenting the results, we first describe the experiment setting and 
methodology that we employed for this placement vulnerability study.

\paragraph{Experiment settings.}
Recall VM placement depends on several placement variables (shown
in~\figref{fig:variables}). We assigned reasonable values to these
placement variables and enumerated through several launch
strategies. A \textit{run} corresponds to one launch strategy and
involves launching multiple VMs from two distinct accounts (i.e.,
subscriptions in Azure and projects in GCE) and checking
for co-residency between all pairs of VMs launched. One account was
designated as a proxy for the victim and the other for the adversary.
We denote a \textit{run configuration} by $v \times a$, where $v$ is the
number of victim instances and $a$ is the number of attacker instances
launched in that run. We varied $v$ and $a$ for all $v$, $a \in \{10,
20, 30\}$ and restricted them to the inequality, $v \le a$, as it increases the
likelihood of achieving co-residency.

Other placement variables that are part of the run configuration
include: victim launch time (including time of the day, day of the
week), delay between victim and attacker VM launches, victim and
attacker instance types and data center location or region where the VMs
are launched. We repeat each run multiple times across all three cloud
providers. The repetition
of experiments is especially required to control the effect of certain
environment variables like time of day. We repeat experiments for each run
configuration over various times of the day and days of the week. We
fix the instance type of VMs to small instances (t2.small on EC2,
g1.small on GCE and small or Standard-A1 on Azure) and data center
regions to us-east for EC2, us-central1-a for GCE and east-us for
Azure, unless otherwise noted. All experiments were conducted
over 3 months between December 2014 to February 2015.

We used a single, local Intel Core i7-2600 machine with 8 SMT cores to
launch VM instances, log instance information and run the co-residency
detection test suite. 

\begin{figure*}[ht!]
\center
\footnotesize
\subfigure[us-central1-a]{
  \begin{tabular}{|c|c|c|c|c|c|c|}
    \hline
    \textbf{Delay}&\textbf{Config.}& \textbf{Mean} & \textbf{S.D.} & \textbf{Min} &\textbf{Median}& \textbf{Max} \\
     (hr.) & & & & & & \\
    \hline
    0 & 10x10 & 0.11 & 0.33 & 0 & 0 & 1 \\
    0 & 10x20 & 0.2 & 0.42 & 0 & 0 & 1 \\
    0 & 10x30 & 0.5 & 0.71 & 0 & 0 & 2 \\
    0 & 20x20 & 0.43 & 0.65 & 0 & 0 & 2 \\
    0 & 20x30 & 1.67 & 1.22 & 0 & 2 & 4 \\
    0 & 30x30 & 1.6 & 1.65 & 0 & 1 & 5 \\
    1 & 10x10 & 0.25 & 0.46 & 0 & 0 & 1 \\
    1 & 10x20 & 0.33 & 0.5 & 0 & 0 & 1 \\
    1 & 10x30 & 1.6 & 1.07 & 0 & 2 & 3 \\
    1 & 20x20 & 1.27 & 1.22 & 0 & 1 & 4 \\
    1 & 20x30 & 2.44 & 1.51 & 0 & 3 & 4 \\
    1 & 30x30 & 3 & 1.12 & 1 & 3 & 5 \\
\hline
\end{tabular}
\label{tab:gce-main-nrcores}
}
 \subfigure[europe-west1-b]{
  \begin{tabular}{|c|c|c|c|c|c|c|}
    \hline
    \textbf{Delay}&\textbf{Config.}& \textbf{Mean} & \textbf{S.D.} & \textbf{Min} &\textbf{Median}& \textbf{Max} \\
     (hr.) & & & & & & \\
    \hline
    0 & 10x10 & 2 & 1.73 & 1 & 1 & 4 \\
    0 & 10x20 & 2.67 & 1.53 & 1 & 3 & 4 \\
    0 & 10x30 & 3 & 2.65 & 1 & 2 & 6 \\
    0 & 20x20 & 3.67 & 1.53 & 2 & 4 & 5 \\
    0 & 20x30 & 2.75 & 2.06 & 0 & 3 & 5 \\
    0 & 30x30 & 12.33 & 2.08 & 10 & 13 & 14 \\
    1 & 10x10 & 2 & 1 & 1 & 2 & 3 \\
    1 & 10x20 & 2 & 1 & 1 & 2 & 3 \\
    1 & 10x30 & 2 & 1.73 & 1 & 1 & 4 \\
    1 & 20x20 & 4.67 & 5.51 & 1 & 2 & 11 \\
    1 & 20x30 & 3.75 & 2.5 & 1 & 3.5 & 7 \\
    1 & 30x30 & 10 & 3 & 7 & 10 & 13 \\
\hline
\end{tabular}
\label{tab:gce-oth-nrcores}
}
\caption{\textbf{Distribution of number of co-resident pairs on GCE.}}
\label{tab:gce-results}
\end{figure*}

\begin{figure*}[ht!]
\centering\
\footnotesize
\subfigure[us-east] {
  \begin{tabular}{|c|c|c|c|c|c|c|}
    \hline
    \textbf{Delay} & \textbf{Config.} & \textbf{Mean} & \textbf{S.D.} & \textbf{Min} & \textbf{Median} & \textbf{Max} \\
     (hr.) & & & & & & \\
    \hline
    0 & $*$ & 0 & 0 & 0 & 0 & 0 \\
    1 & 10x10 & 0.44 & 0.73 & 0 & 0 & 2 \\
    1 & 10x20 & 1.11 & 1.17 & 0 & 1 & 3 \\
    1 & 10x30 & 1.4 & 1.43 & 0 & 1.5 & 4 \\
    1 & 20x20 & 3.57 & 2.59 & 0 & 3.5 & 9 \\
    1 & 20x30 & 3.78 & 1.79 & 1 & 4 & 7 \\
    1 & 30x30 & 3.89 & 2.09 & 2 & 3 & 9 \\
    \hline
  \end{tabular}
\label{tab:ec2-main-nrcores}
}
\subfigure[us-west-1 (CA)] {
  \begin{tabular}{|c|c|c|c|c|c|c|}
    \hline
    \textbf{Delay} & \textbf{Config.} & \textbf{Mean} & \textbf{S.D.} & \textbf{Min} & \textbf{Median} & \textbf{Max} \\
     (hr.) & & & & & & \\
    \hline
    0 & $*$ & 0 & 0 & 0 & 0 & 0 \\
    0 & 20x20 & 10.33 & 8.96 & 0 & 15 & 16 \\
    1 & 10x10 & 1.67 & 0.58 & 1 & 2 & 2 \\
    1 & 10x20 & 2.33 & 0.58 & 2 & 2 & 3 \\
    1 & 10x30 & 5.33 & 2.52 & 3 & 5 & 8 \\
    1 & 20x20 & 8.33 & 4.51 & 4 & 8 & 13 \\
    1 & 20x30 & 5.5 & 3.87 & 2 & 4.5 & 11 \\
    1 & 30x30 & 8.33 & 6.66 & 4 & 5 & 16 \\
    \hline
  \end{tabular}
\label{tab:ec2-oth-nrcores}
}
\caption{\textbf{Distribution of number of co-resident pairs on
EC2.}}
\label{tab:ec2-results}
\end{figure*}

\begin{figure}[ht!]
\centering\
\footnotesize
  \begin{tabular}{|c|c|c|c|c|c|c|}
    \hline
    \textbf{Delay} (hr.) & \textbf{Config.} & \textbf{Mean} & \textbf{S.D.} & \textbf{Min} & \textbf{Median} & \textbf{Max} \\
    \hline
    0 & 10x10 & 15.22 & 19.51 & 0 & 14 & 64 \\
    0 & 10x20 & 3.78 & 4.71 & 0 & 3 & 14 \\
    0 & 10x30 & 4.25 & 6.41 & 0 & 2.5 & 19 \\
    0 & 20x20 & 9.67 & 8.43 & 0 & 8 & 27 \\
    0 & 20x30 & 2.38 & 1.51 & 1 & 2 & 5 \\
    0 & 30x30 & 24.57 & 36.54 & 1 & 6 & 99 \\
    1 & 10x10 & 2.78 & 3.87 & 0 & 1 & 12 \\
    1 & 10x20 & 0.78 & 1.2 & 0 & 0 & 3 \\
    1 & 10x30 & 0.75 & 1.39 & 0 & 0 & 3 \\
    1 & 20x20 & 0.67 & 1.66 & 0 & 0 & 5 \\
    1 & 20x30 & 0.86 & 0.9 & 0 & 1 & 2 \\
    1 & 30x30 & 4.71 & 9.89 & 0 & 1 & 27 \\
    \hline
  \end{tabular}
\caption{\textbf{Distribution of number of co-resident pairs on
Azure.} Region: East US 1.}
\label{tab:azure-nrcores}
\end{figure}

\paragraph{Implementation and the Cloud APIs.}
In order to automate our experiments, we used Python and the
libcloud\footnote{We used libcloud version 0.15.1 for EC2, and
a modified version of 0.16.0 for GCE to support the use of multiple
accounts in GCE.} library~\cite{libcloud} to interface with EC2 and
GCE. Unfortunately, libcloud did not support Azure. The only Azure
cloud API on Linux platform was a node.js library and a cross-platform
command-line interface (CLI). We built a wrapper around the CLI. There
were no significant differences across different cloud APIs except
that Azure did not have any explicit interface to launch multiple VMs
simultaneously.

As mentioned in the experiment settings, we experimented with various
delays between the victim and attacker VM launches (0, 1, 2, 4 \ldots
hours). To save money, we reused the same set of victim instances for
each of the longer runs. That is, for the run configuration of 10x10
with 0, 1, 2, and 4 hours of delay between victim and attacker VM
launches, we launched the victim VMs only once at the start of the
experiment. After running co-residency tests on the first set of VM
pairs, we terminated all the attacker instances and relaunched
attacker VM instances after appropriate delays (say 1 hour) and rerun
the tests with {\em the same set} of victim VMs. We repeat this until
we experiment with all delays for this configuration.
We call this methodology the {\em leap-frog method}. It is also
important to note that zero delay here means parallel launch of VMs
from our test machine (and not sequential launch of VMs from one
account after another), unless otherwise noted.

In the sections below, we take a closer look at the effect of varying
one placement variable while keeping other variables fixed across
all the cloud providers. In each case, we use three metrics to measure
the degree of co-residency: chances of getting at least one
co-resident instance across a number of runs (or success rate), average number of
co-resident instances over multiple runs and average coverage
(i.e., fraction of victim VMs with which attacker VMs were
co-resident). Although these experiments were done with victim VMs
under our control, the results can be extrapolated to guide an
attacker's launch strategy for an uncooperative victim. We also
discuss a set of such strategic questions that the results help
answer. At the end of this section, we summarize and calculate the cost of
several interesting launch strategies
and evaluate the public clouds against our reference
placement policy.

\subsection{Effect of Number of Instances}
In this section, we observe the placement behavior while varying the
number of victim and attacker instances. Intuitively, we expect the
chances of co-residency to increase with the number of
attacker and victim instances.

\begin{figure}[ht!]
\center
\includegraphics[scale=0.75]{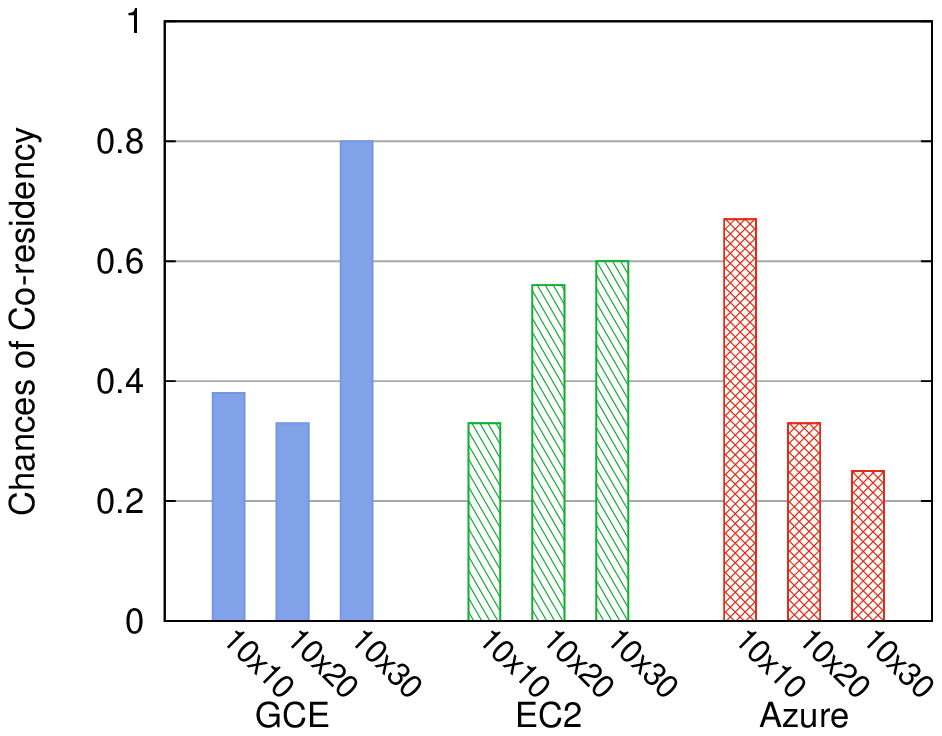}
\caption{{\bf Chances of co-residency of 10 victim instances with
varying number of attacker instances.} All these results are from one
data center region (EC2: us-east, GCE: us-central1-a, Azure: East US)
and the delays between victim and attacker instance launch were 1
hour. Results are over at least 9 runs per run configuration with at
least 3 runs per time of day.}
\label{fig:vary-A-chance}
\end{figure}

\paragraph{Varying number of attacker instances.}
Keeping all the placement variables constant including the number of
victim instances, we measure the chance of co-residency over multiple
runs. The result of this experiment helps to answer the question:
How many VMs should an adversary launch to increase the chance of
co-residency?

As is shown in \figref{fig:vary-A-chance}, the placement behavior
changes across different cloud providers. For GCE and EC2, we observe
that higher the number of attacker instances relative to the victim
instances, the higher the chance of co-residency is.
\figref{tab:gce-main-nrcores} and~\ref{tab:ec2-main-nrcores} show the
distribution of number of co-resident VM pairs on GCE and EC2,
respectively. The number of co-resident VM pairs also increases with
the number of attacker instances, implying that the coverage of an
attack could be increased with larger fraction of attacker instances
than the target VM instances if the launch times are coordinated.

Contrary to our expectations, the placement behavior observed on Azure
is the inverse. The chances of co-residency with 10 attacker instances
are almost twice as high as with 30 attacker instances. This is also reflected
in the distribution of number of co-residency VM pairs (shown
in~\figref{tab:azure-nrcores}). Further investigation revealed a
correlation between the number of victim and attacker instances
launched and the chance of co-residency. That is, for the run
configuration of 10x10, 20x20 and 30x30, where number of victim and
attacker instances are the same, and with 0 delay, the chance of
co-residency were equally high for all these configurations
(between 0.9 to 1). This suggests a possible placement policy that
collates VM launch requests together based on their request size and
places them on the same group of machines.

\begin{figure}[ht!]
\center
\includegraphics[scale=0.75]{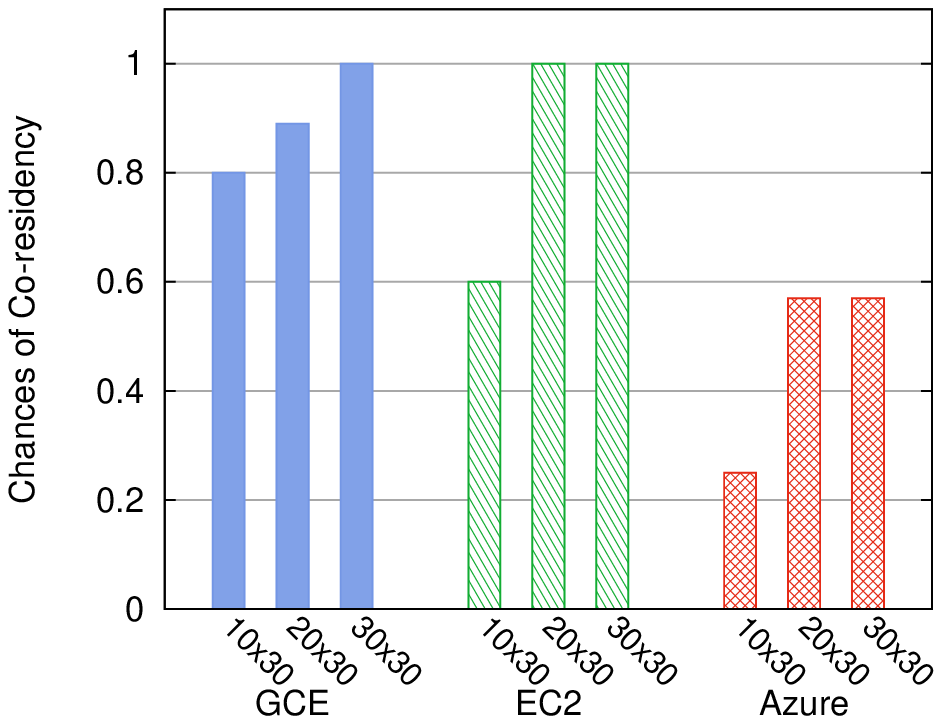}
\caption{{\bf Chances of co-residency of 30 attacker instances with
varying number of victim instances.} All these
results are from one data center region (EC2: us-east, GCE:
us-central1-a, Azure: East US) and the delays between victim and
attacker instance launch were 1 hour. Results are over at least 9 runs
per run configuration with at least 3 runs per time of day.}
\label{fig:vary-V-chance}

\end{figure}

\paragraph{Varying number of victim instances.}
Similarly, we also varied the number of victim instances by keeping the number
of attacker instances and other placement variables constant (results shown
in~\figref{fig:vary-V-chance}).  We expect the chance of co-residency to
increase with the number of victims targeted. Hence, the results presented here
help an adversary answer the question: What are the chances of co-residency with
varying sizes of target victims?

As expected, we see an increase in the chances of co-residency with increasing
number of victim VMs across all cloud providers. We see that the absolute value
of the chance of co-residency is lower for Azure than other clouds. This may be
the result of significant additional delay between victim and attacker launch
times in Azure as a result of our methodology (more on this later).

\subsection{Effect of Instance Launch Time}
In this section, we answer two questions that aid an adversary to
design better launch strategies: How quickly should an attacker launch VMs
after the victim VMs are launched? Is there any increase in chance
associated with the time of day of the launch?


\begin{figure}
\center
\includegraphics[scale=0.75]{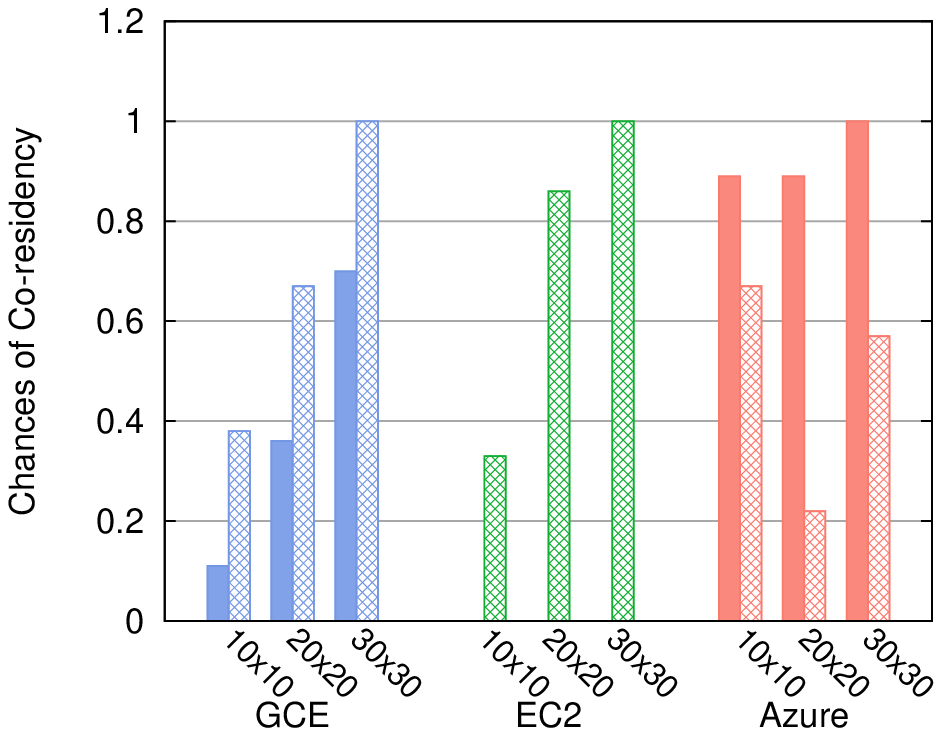}
\caption{{\bf Chances of co-residency with varying delays between
victim and attacker launches.} \textit{Solid} boxes
  correspond to zero delay (simultaneous launches) and
  \textit{gauze}-like boxes correspond to 1 hour delay between victim
  and attacker launches. We did not observe any co-resident
  instances for runs with zero delay on EC2. All these results are
from one data center region (EC2: us-east, GCE: us-central1-a, Azure:
  East US). Results are over at least 9 runs per run configuration with
  at least 3 runs per time of day.}
    \label{fig:vary-delay-chance}

\end{figure}

\paragraph{Varying delay between attacker and victim launches.}
The result of varying the delay between 0 (i.e., parallel launch) and 1 hour
delay is shown in~\figref{fig:vary-delay-chance}. We can make two immediate
observations from this result.

\begin{figure}
\centering\
\footnotesize
  \begin{tabular}{|c|c|c|c|c|c|c|c|}
    \hline
    \textbf{Delay} & \textbf{Mean} & \textbf{S.D.} & \textbf{Min} & \textbf{Median} & \textbf{Max} & \textbf{Success} \\
    & & & & & & \textbf{rate} \\
    \hline
    0+ & 0.6 & 1.07 & 0 & 0 & 3 & 0.30 \\
    5 min & 1.38 & 0.92 & 0 & 1 & 3 & 0.88 \\
    1 hr & 3.57 & 2.59 & 0 & 3.5 & 9 & 0.86 \\
    \hline
  \end{tabular}
  \caption{\textbf{Distribution of number of co-resident pairs and success rate
      or chances of co-residency
      for shorter delays under 20x20 run configuration in EC2.} A delay with 0+
    means victim and attacker instances were launched sequentially,
    i.e. attacker instances were not launched until all victim instances were
    running. The results averaged are over 9 runs with 3 runs per time of day.}
\label{tab:ec2-short-delay}
\end{figure}

The first observation reveals a significant artifact of EC2's placement
policy: VMs launched within a short time window are never co-resident
on the same machine. This observation helps an adversary to avoid
such a strategy. We further
investigated placement behaviors on EC2 with shorter non-zero delays
in order to find the duration of this time window in which there are
zero co-residency (results shown in
~\figref{tab:ec2-short-delay}). We found that this time window is very short and
that even a sequential launch of instances (denoted by 0+)
could result in co-residency.

The second observation shows that non-zero delay on GCE and zero delay
on Azure increases the chance of co-residency and hence directly
benefits an attacker. It should be noted that on
Azure, the launch delays between victim and attacker instances were
longer than 1 hour due to our leap-frog experimental
methodology; the actual delays between the VM launches were, on
average, 3 hours (with a maximum delay of 10 hours for few runs).
This higher delay was more common in runs with larger number of
instances as there were significantly more false positives, which
required a separate sequential phase to resolve (see
\secref{ssec:coop-test}).

\begin{figure}
\center
\includegraphics[scale=0.75]{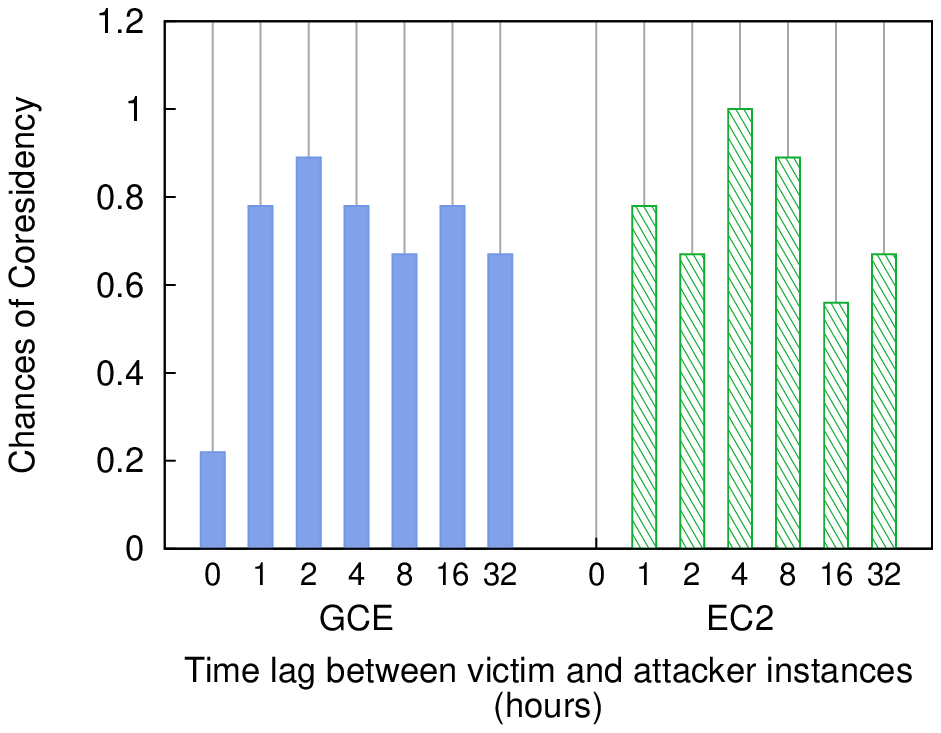}
\caption{{\bf Chances of co-residency over long periods.} Results include 9 runs over two
  weeks with 3 runs per time of day under 20x20 run configuration. Note that we only conducted 3 runs for 32
  hour delay as opposed to 9 runs for all other delays.}
\label{fig:long-running}
\end{figure}

We also experimented with longer delays on EC2 and
GCE to understand whether and how quickly the
chance of co-residency drops with increasing delay (results shown
in~\figref{fig:long-running}). Contrary to our expectation, we did not
find the chance of co-residency to drop to zero even for delays as high
as 16 and 32 hours. We speculate that the
reason for this observation could be that the system was under constant churn where some neighboring VMs on the
victim's machine were terminated. Note that our leap-frog methodology may,
in theory, interfere with the VM placement. But it is noteworthy that we observed
increased number of unique co-resident pairs with increasing delays, suggesting
fresh co-residency with victim VMs over longer delays. 


\begin{figure}
\centering\
\footnotesize
  \begin{tabular}{|c|c|c|c|}

    \hline
    \multicolumn{4}{|c|}{\bf Chances of Co-residency} \\
    \hline
    \textbf{Cloud} & \textbf{Morning} & \textbf{Afternoon} & \textbf{Night}\\
     & 02:00 - 10:00 & 10:00 - 18:00 & 18:00 - 02:00\\  
    \hline
    GCE & 0.68 & 0.61& 0.78 \\
    EC2 & 0.89 & 0.73 & 0.6  \\
    \hline

  \end{tabular}


\caption{\textbf{Effect of time of day.} Chances of
  co-residency when an attacker changes the launch time of his instances.
The results were aggregated across all run configurations with 1 hour
delay between victim and attacker launch times. All times are in PT.}
\label{tab:tod-effect} 
\end{figure}

\paragraph{Effect of time of day.}
Prior works have shown that churn or load is often correlated
with the time of day~\cite{who-was-paper}. Our simple reference
placement policy does not have a notion of load and hence have no
effect on time of day. In reality, with limited number of servers in
datacenters and limited number of capacity per host, load on the system
has direct effect on the placement behavior of any placement policy.

As expected, we observe small effect on VM placement based on the time
of day when attacker instances are launched (results shown
in~\figref{tab:tod-effect}). Specifically, there is a slightly higher
chance of co-residency if the attacker instances are launched in the
early morning for EC2 and at night for GCE.

\begin{figure}
\center
    \subfigure[GCE]{
      \includegraphics[scale=0.75]{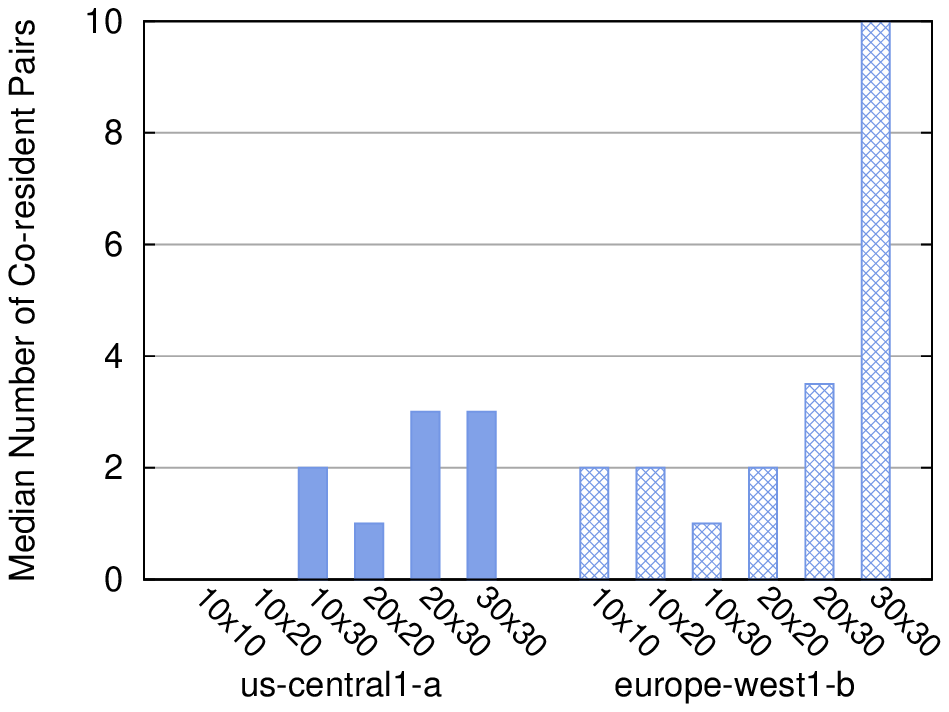}
    }
    \subfigure[EC2]{
      \includegraphics[scale=0.75]{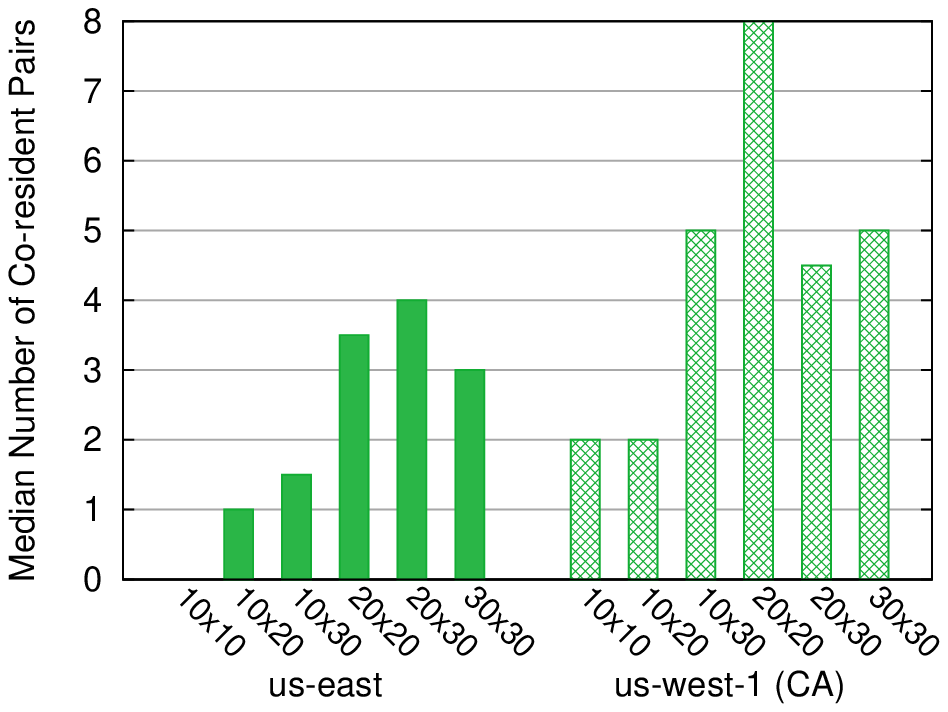}
    }
    \vspace{-0.15in}
    \caption{{\bf Median number of co-resident pairs across two
        regions.} The box plot shows the median number of co-resident
      pairs excluding co-residency within the same account. Results
      are over at least 3 run per run configuration (x-axis).}
    \label{fig:oth-nr-cores}
\end{figure}

\subsection{Effect of Data Center Location}
All the above experiments were conducted on relatively popular regions in each
cloud (especially true for EC2~\cite{who-was-paper}). In this section, we report
the results on other smaller and less popular regions.  As the regions are less
popular and have relatively fewer machines, we expect higher co-residency rates
and more co-resident instances. \figref{fig:oth-nr-cores} shows the median
number of co-resident VM pairs placed in these regions alongside the results
for popular regions. The distribution of number of co-resident instances is 
shown in~\figref{tab:gce-oth-nrcores} and~\ref{tab:ec2-oth-nrcores}.

The main observation from these experiments is that there is a higher
chance of co-residency in these smaller regions than the larger, more
popular regions. Note that we
placed at least one co-resident pair in all the runs in these
regions. Also the higher number of co-resident pairs also suggests a
larger coverage over victim VMs in these smaller regions.

One anomaly that we found during two 20x20 runs on EC2 between
$30^{th}$ and $31^{st}$ of January 2015, when we observed an unusually
large number of co-resident instances (including three VMs from the
same account). We believe this anomaly may be a result of an internal
management incident in the Amazon EC2 us-west-1 region.

\begin{figure}
\center
\includegraphics[scale=0.75]{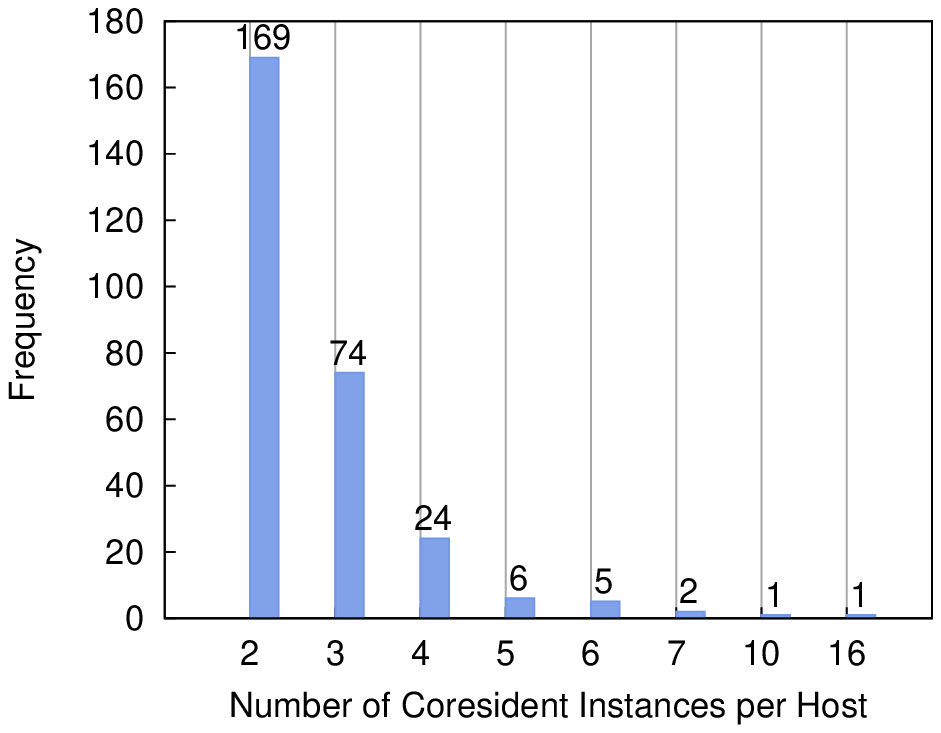}
\caption{{\bf Distribution of number of co-resident instances per host on
    Azure.} The results shown are across all the runs. We saw at most 2
  instances per host in EC2 and at most 3 instances per host in GCE.}
\label{fig:azure-cores}

\end{figure}

\subsection{Other Observations}

We report several other interesting observations in this section.
First, we found more than two VMs can be co-resident on the same host
on both Azure and GCE, but not on EC2. \figref{fig:azure-cores} shows the
distribution of number of co-resident instances per host.
Particularly, in one of the runs, we placed 16 VMs on a single
host.

Another interesting observation is related to co-resident instances
from the same account. We term them as {\em self-co-resident instances}.
We observed many self-co-resident pairs on GCE and Azure
(not shown). On the other hand, we never noticed any self co-resident
pair on EC2 except for the anomaly in us-west-1. Although we did not
notice any effect on the actual chance of co-residence, we believe
such placement behaviors (or the lack of) may affect VM placement.

We also experimented with medium instances and successfully placed few
co-located VMs on both EC2 and GCE by employing similar successful strategies
learned with small instances.



\subsection{Cost of Launch Strategies}
\begin{figure}
\center
\includegraphics[scale=0.75]{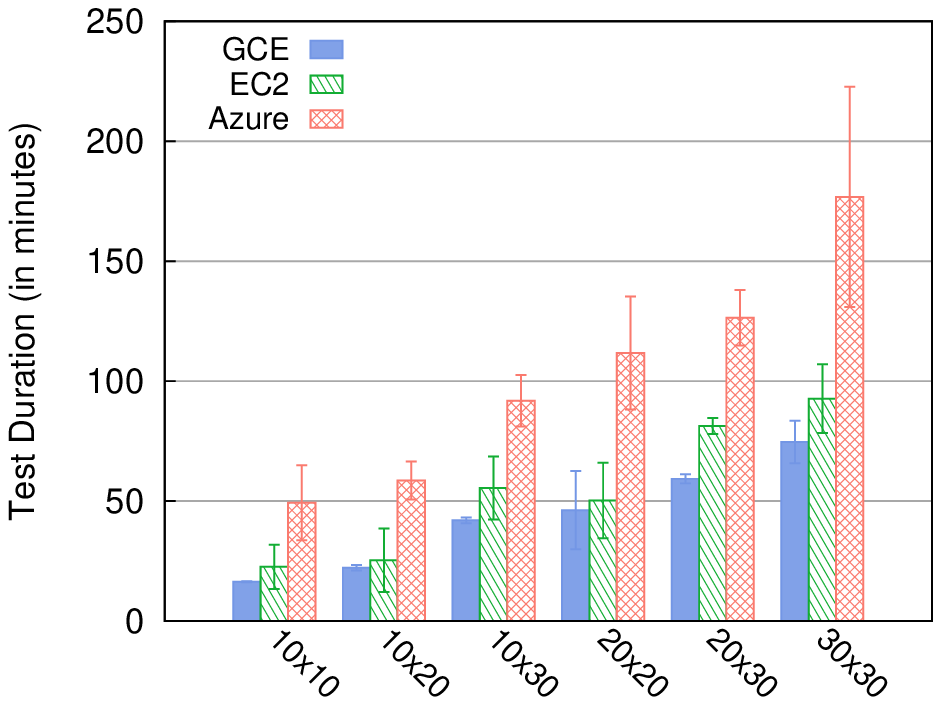}
\vspace{-0.2in}
\caption{{\bf Launch strategy and co-residency detection execution
    times.} The run configurations $v \times a$ indicates the number of
  victims vs. number of attackers launched. The error bars show the
  standard deviation across at least 7 runs.}
    \label{fig:test-duration}
\end{figure}

\begin{figure}
  \center
  \footnotesize
  \begin{tabular}{|c|c|c|c|c|c|c|}
    \hline
    {\bf Run} & \multicolumn{3}{c|}{\bf Average Cost (\$)} & \multicolumn{3}{c|}{\bf Maximum Cost (\$)} \\
    {\bf config.} & GCE & EC2 & Azure & GCE & EC2 & Azure\\
    \hline
    10x10 & \textbf{0.137} & 0.260 & 0.494 & 0.140 & 0.260 & 0.819\\
    10x20 & 0.370 & 0.520 & 1.171 & 0.412 & 0.520 & 1.358\\
    10x30 & 1.049 & 0.780 & 2.754 & 1.088 & 1.560 & 3.257\\
    20x20 & 0.770 & 0.520 & 2.235 & 1.595 & 1.040 & 3.255\\
    20x30 & 1.482 & 1.560 & 3.792 & 1.581 & 1.560 & 4.420\\
    30x30 & 1.866 & 1.560 & 5.304 & 2.433 & 1.560 & \textbf{7.965}\\
    \hline
  \end{tabular}
  \caption{\textbf{Cost of running a launch strategy} (in
    dollars). Maximum cost column refers to the maximum cost we
    incurred out of all the runs for that particular configuration and
    cloud provider. The cost per hour of small
    instances at the time of this study were: 0.05, 0.026 and 0.06
    dollars for GCE, EC2 and Azure, respectively. The minimum and
    maximum costs are in \textbf{bold}.}
\label{tab:cost-strategies}
\end{figure}

Recall that the cost of a launch strategy from
\secref{sec:threat-model}, $C_{S} = a * P(a_{type}) * T_{d}(v,a)$. In
order to calculate this cost, we need $T_{d}(v,a)$ which is the time
taken to detect co-location with $a$ attackers and $v$
victims. \figref{fig:test-duration} shows the average time taken to
complete launching attacker instances and complete co-residency detection
for each run configuration. Here the measured co-residency detection
is the parallelized version discussed in \secref{ssec:coop-test} and
also includes time taken to detect co-residency within each tenant
account. Hence, for these reasons the time to detect co-location is an
upper bound for a realistic and highly optimized co-residency
detection mechanism. 

We calculate the cost of executing each launch strategy under the
three public clouds. The result is summarized
in~\figref{tab:cost-strategies}. Note that we only consider the cost
incurred by the compute instances because the cost for
other resources such as network and storage, was insignificant. Also note
that EC2 bills every hour even if an instance runs less than an
hour~\cite{ec2-pricing}, whereas GCE and Azure charge per minute of
instance activity. This difference is considered in our cost
calculation. Overall, the maximum cost
we incurred was about \$8 for running 30 VMs for 4 hours 25 minutes on
Azure and a minimum of 14 cents on GCE for running 10 VMs for 17
minutes. We incurred the highest cost for all the launch strategies in
Azure because of overall higher cost per hour and partly due to longer tests due to our
co-residency detection methodology. 

\subsection{Summary of Placement Vulnerabilities}


\noindent In this section, we return to the secure reference placement policy
introduced in \secref{sec:threat-model} and use it to identify placement
vulnerabilities across all the three clouds. Recall that the probability of at
least one pair of co-residency under this random placement policy is given by
$\Prob{\E_{a}^{v} > 0} = 1 - (1 - v/N)^a$, where $\E_{a}^{v}$ is the random
variable denoting the number of co-location observed when placing $a$ attacker
VMs among $N=1000$ total machines where $v$ machines are already picked for the
$v$ victim VMs. First, we evaluate this
probability for various run configurations that we experimented with in the
public clouds. The probabilities are shown in~\figref{fig:ref-mode-prob}.


\begin{figure}
  \center
  \footnotesize
    \begin{tabular*}{\columnwidth}{c|cccccc}
    \hline
    \textbf{Run Config.} & 10x10 & 10x20 & 10x30 & 20x20 & 20x30 &
30x30 \\\hline
$ \Prob{\E_{a}^{v} > 0}$ & 0.10 & 0.18 & 0.26 & 0.33 & 0.45 & 0.60 \\
    \hline
  \end{tabular*}
  \caption{\textbf{Probability of co-residency under the reference
      placement policy.}}
  \label{fig:ref-mode-prob}
\end{figure}

\begin{figure}
  \center
  \footnotesize
    \begin{tabular}{|c|c|c|c|c|c|}
    \hline
    \textbf{Strategy} & $v$ \& $a$ & $a'$ & \textbf{Cost benefit} (\$) & \textbf{Normalized}\\
     & & & & \textbf{success}\\
    \hline
    S1 \& S2 & 10 & 688 & 113.87 & 10\\
    S3 & 30 & 227 & 32.75 & 1.67\\
    S4(i) & 20 & 105 & 4.36 & 2.67\\
    S4(ii) & 20 & 342 & 53.76 & 3.03\\
    S5 & 20 & 110 & 4.83 & 1.48\\
    \hline
  \end{tabular}
  \caption{\textbf{Cost benefit analysis.} $N = 1000$, $P(a_{type}) = 0.026$,
    which is the cost per instance-hour on EC2 (the cheapest). For
    simplicity $T_{d}(v,a) = (v * a) * 3.85$, where $3.85$ is fastest average
    time to detect co-residency per instance-pair. Here, $v \times a$ is the run
    configuration of the strategy under test. Note that the cost benefit is the
    additional cost incurred under the reference policy, hence is equal to cost
    incurred by $a'-a$ additional VMs.}
  \label{fig:cb-analysis}  
\end{figure}

Recall that a launch strategy in a cloud implies a placement vulnerability in
that cloud's placement policy if its normalized
success rate is greater than 1. The normalized success rate of the strategy
is the ratio of the chance of co-location under that launch strategy to the probability of
co-location in the reference policy ($\Prob{\E_{a}^{v} > 0}$). Below is a list
of selected launch strategies that escalate to placement vulnerabilities using
our reference policy with their normalized success rate in parenthesis.

\begin{packeditemize}

\item[\textit{(S1)}] In Azure, launch ten attacker VMs closely after the
  victim VMs are launched (1.0/0.10).

\item[\textit{(S2)}] In EC2 and GCE, if there are known victims in
  any of the smaller datacenters, launch at least ten attacker VMs
  with a non-zero delay (1.0/0.10).

\item[\textit{(S3)}] In all three clouds, launch 30 attacker instances, either
  with no delay (Azure) or one hour delay (EC2, GCE) from victim launch, to get
  co-located with one of the 30 victim instances (1.00/0.60).

\item[\textit{(S4)}] (i) In Amazon EC2, launch 20 attacker VMs with
  a delay of 5 minutes or more after the victims are launched
  (0.88/0.33). (ii) The optimal delay between
  victim and attacker VM launches is around 4 hours for a 20x20 run (1.00/0.33).

\item[\textit{(S5)}] In Amazon EC2, launch the attacker VMs with
  1 hour after the victim VMs are launched where the time of day falls
  in the early morning, i.e., 02:00 to 10:00hrs PST 
  (0.89/0.60).
\end{packeditemize}

\paragraph{Cost benefit.} 
Next, we quantify the cost benefit of each of these strategies over the
reference policy. As the success rate of any launch strategy on a vulnerable
placement policy is greater than what is possible in the reference policy, 
we need more attacker instances in
the reference policy to achieve the same success rate. We
calculate this number of attacker instances $a'$ using: $ a' = \ln(1 -
S_{a}^{v})/\ln(1 - {v}/{N})$, where, $S_{a}^{v}$ is the success rate of a
strategy with run configuration of $v \times a$. The result of this calculation is
presented in~\figref{fig:cb-analysis}. The result shows that the best strategy,
S1 and S2, on all three cloud providers is \$114 cheaper than what is possible
in the reference policy. 

It is also evident that these metrics enable evaluating and comparing various
launch strategies and their efficacy on various placement policies both
on robust placements and attack cost. For example, note that although the
normalized success rate of {\em S3} is lower than {\em S4}, it has a higher cost
benefit for the attacker. 

%

\subsection{Limitations}

Although we exhaustively experimented with a variety of
placement variables, the results have limitations. One major
limitation of this study is the number of placement variables and the
set of values for the variables that we used to experiment. For
example, we limited our experiments with only one instance type, one
availability zone per region and used only one account for the victim
VMs. Although different instance types may exhibit different placement
behavior, the presented results hold strong for the chosen
instance type. The
only caveat that may affect the results is if the placement policy uses
account ID for VM placement decisions. Since, we experimented
with only one victim account (separate from the designated attacker
account) across all providers, these results, in the worst case, may
have captured the placement behavior of an unlucky victim account that
was subject to similar placement decisions (and hence co-resident) as
that of the VMs from the designated attacker account. 

Even though we ran at least 190 runs per cloud provider over a period
of 3 months to increase statistical significant of our results, we
were still limited to at most 9 runs per run configuration (with 3
runs per time of day).  These limitations have only minor bearing on
the results presented, if any, and the reported results are
significant and impactful for cloud computing security research.

\section{Related Work}
\label{sec:relwork}


\vspace{-0.1in}
\paragraph{VM placement vulnerability studies.} Ristenpart et
al.~\cite{rist:hey-you:ccs:2009} first studied the placement
vulnerability in public clouds, which showed that a malicious cloud
tenant could place one of his VMs on the same machine as a target VM
with high probability. Placement vulnerabilities exploited in their
study include publicly available mapping of VM's public/internal IP
addresses, disclosure of Dom0 IP addresses, and a shortcut
communication path between co-resident VMs.  Their study was followed
by Xu et al.~\cite{xu2011exploration} and further extended by Herzberg
et al.~\cite{Herzberg:2013:CSD}. However, the results of these studies
have been outdated by the recent development of cloud technologies,
which is the main motivation of our work. 

Concurrent with our work, Xu et al.~\cite{XuWaWu:2015} conducted a systematic
measurement study of co-resident threats in Amazon EC2. Their focus, however, is
in-depth evaluation of co-residency detection using network route traces and
quantification of co-residence threats on older generation instances with EC2's
classic networking (prior to Amazon VPC). In contrast, we study placement
vulnerabilities in the context of VPC on EC2, as well as on Azure and GCE. The
two studies are mostly complementary and strengthen the arguments made by each
other.

New VM placement policies to defend against placement attacks have been
studied by Han et al.~\cite{HaAlCh:13, HaChAl:14} and Azar et
al.~\cite{Azar:2014:CC}. It is unclear, however, whether their proposed
policies work against the performance and reliability goals of public cloud providers.

\paragraph{Co-residency detection techniques.}
Techniques for co-residency detection have been studied in various
contexts. We categorize these techniques into one of the two classes:
side-channel approaches to detecting co-residency with
\textit{uncooperative} VMs and covert-channel approaches to detecting
co-residency with \textit{cooperative} VMs.

Side-channels allow one party to exfiltrate secret information from
another; therefore these approaches may be adapted in practical
placement attack scenarios with targets not controlled by the
attackers. Network round-trip timing side-channel was used by
Ristenpart et al.~\cite{rist:hey-you:ccs:2009} to detect co-residency.
Zhang et al.~\cite{zhang11:homealone} developed a system called
\emph{HomeAlone} to enable VMs to detect third-party VMs using timing
side-channels in the last level caches. Bates et
al.~\cite{bates2012detecting} proposed a side-channel for co-residency
detection by causing network traffic congestion in the host NICs from
attacker-controlled VMs; the interference of target VM's performance,
if the two VMs are co-resident, should be detectable by remote
clients. Kohno et al.~\cite{kohno2005remote} explored techniques to
fingerprint remote machines using timestamps in TCP or ICMP based
network probes, although their approach was not designed for
co-residency detection.  However, none of these approaches works
effectively in modern cloud infrastructures.   


Covert-channels on shared hardware components can be used for
co-residency detection when both VMs under test are cooperative.
Coarse-grained covert-channels in CPU caches and hard disk drives were
used in Ristenpart et al.~\cite{rist:hey-you:ccs:2009} for
co-residency confirmation.  Xu et al.~\cite{xu2011exploration}
established covert-channels in shared last level caches between two
colluding VMs in the public clouds.  Wu el al.~\cite{wu2012whispers}
exploited memory bus as a covert-channel on modern x86 processors, in
which the sender issues atomic operations on memory blocks spanning
multiple cache lines to cause memory bus locking or similar effects on
recent processors. However, covert-channels proposed in the latter two
studies were not designed for co-residency detection, while those
developed in our work are tuned for this purpose.




\section{Conclusion and Future Work}
\label{sec:conclude}
Multi-tenancy in public clouds enable co-residency attacks. In this paper, we
revisited the problem of placement --- can an attacker achieve co-location? ---
in modern public clouds. We find that while past techniques for verifying
co-location no longer work, insufficient performance isolation in hardware still
allows detection of co-location. Furthermore, we show that in the three popular
cloud providers (EC2, GCE and Azure), achieving co-location is surprisingly
simple and cheap. It is even simpler and costs nothing to achieve
co-location in some PaaS clouds. Our results demonstrate that even though cloud
providers have massive datacenters with numerous physical servers, the
chances of co-location are far higher than expected. More work is needed to
achieve a better balance of efficiency and security using smarter 
co-location-aware placement policies.

\section*{Acknowledgments}

This work was funded by the National Science Foundation under grants CNS-1330308,
CNS-1546033 and CNS-1065134. Swift has a significant financial interest in Microsoft Corp.

\bibliographystyle{abbrv}
\setlength{\itemsep}{-1.5pt}
{\footnotesize \bibliography{placement.bib}}

\begin{thebibliography}{10}

\bibitem{ec2-instance-store}
Amazon ec2 instance store.
\newblock
  \url{http://docs.aws.amazon.com/AWSEC2/latest/UserGuide/InstanceStorage.html}.

\bibitem{amazon:ec2}
{Amazon} elastic compute cloud.
\newblock \url{http://aws.amazon.com/ec2/}.

\bibitem{libcloud}
Apache libcloud.
\newblock \url{http://libcloud.apache.org/}.

\bibitem{aws-elastic-bs}
Aws elastic beanstalk.
\newblock \url{http://aws.amazon.com/elasticbeanstalk/}.

\bibitem{aws-reinvent-2014}
Aws innovation at scale, re:invent 2014, slide 9-10.
\newblock
  \url{http://www.slideshare.net/AmazonWebServices/spot301-aws-innovation-at-scale-aws-reinvent-2014}.

\bibitem{gce}
Google compute engine.
\newblock \url{https://cloud.google.com/compute/}.

\bibitem{gce-disks}
Google compute engine -- disks.
\newblock \url{https://cloud.google.com/compute/docs/disks/}.

\bibitem{gce-autoscaler}
Google compute enginer autoscaler.
\newblock \url{http://cloud.google.com/compute/docs/autoscaler/}.

\bibitem{haproxy}
Haproxy: The reliable, high performance tcp/http load balancer.
\newblock \url{http://www.haproxy.org/}.

\bibitem{heroku-network}
Heroku devcenter: Dynos and the dyno manager, ip addresses.
\newblock \url{https://devcenter.heroku.com/articles/dynos#ip-addresses}.

\bibitem{heroku}
Heroku {PaaS} system.
\newblock \url{https://www.heroku.com/}.

\bibitem{ivy-bridge-llc}
{Intel Ivy Bridge cache replacement policy}.
\newblock \url{http://blog.stuffedcow.net/2013/01/ivb-cache-replacement/}.

\bibitem{olio-workload}
Olio workload.
\newblock \url{https://cwiki.apache.org/confluence/display/OLIO/The+Workload}.

\bibitem{rightscale}
Rightscale.
\newblock \url{http://www.rightscale.com}.

\bibitem{azure-disks}
Virtual machine and cloud service sizes for azure.
\newblock \url{https://msdn.microsoft.com/en-us/library/azure/dn197896.aspx}.

\bibitem{azure}
Windows azure.
\newblock \url{http://www.windowsazure.com/}.

\bibitem{ec2-pricing}
Amazon ec2 pricing, 2015.
\newblock \url{http://aws.amazon.com/ec2/pricing/}.

\bibitem{aws-vpc-whitepaper}
{Amazon Web Services}.
\newblock Extend your it infrastructure with amazon virtual private cloud.
\newblock Technical report, Amazon, 2013.

\bibitem{Azar:2014:CC}
Y.~Azar, S.~Kamara, I.~Menache, M.~Raykova, and B.~Shepard.
\newblock Co-location-resistant clouds.
\newblock In {\em In Proceedings of the ACM Workshop on Cloud Computing
  Security}, pages 9--20, 2014.

\bibitem{bates2012detecting}
A.~Bates, B.~Mood, J.~Pletcher, H.~Pruse, M.~Valafar, and K.~Butler.
\newblock Detecting co-residency with active traffic analysis techniques.
\newblock In {\em Proceedings of the 2012 ACM Workshop on Cloud Computing
  Security Workshop}, pages 1--12. ACM, 2012.

\bibitem{beach2014vpc}
B.~Beach.
\newblock Virtual private cloud.
\newblock In {\em Pro Powershell for Amazon Web Services}, pages 67--88.
  Springer, 2014.

\bibitem{money-socc12}
B.~Farley, A.~Juels, V.~Varadarajan, T.~Ristenpart, K.~D. Bowers, and M.~M.
  Swift.
\newblock More for your money: Exploiting performance heterogeneity in public
  clouds.
\newblock In {\em Proceedings of the Third ACM Symposium on Cloud Computing}.
  ACM, 2012.

\bibitem{cloudsuite2}
M.~Ferdman, A.~Adileh, O.~Kocberber, S.~Volos, M.~Alisafaee, D.~Jevdjic,
  C.~Kaynak, A.~D. Popescu, A.~Ailamaki, and B.~Falsafi.
\newblock Clearing the clouds: a study of emerging scale-out workloads on
  modern hardware.
\newblock In {\em Proceedings of the seventeenth international conference on
  Architectural Support for Programming Languages and Operating Systems}. ACM,
  2012.

\bibitem{HaAlCh:13}
Y.~Han, T.~Alpcan, J.~Chan, and C.~Leckie.
\newblock Security games for virtual machine allocation in cloud computing.
\newblock In {\em Decision and Game Theory for Security}. Springer
  International Publishing, 2013.

\bibitem{HaChAl:14}
Y.~Han, J.~Chan, T.~Alpcan, and C.~Leckie.
\newblock Virtual machine allocation policies against co-resident attacks in
  cloud computing.
\newblock In {\em IEEE International Conference on Communications,}, 2014.

\bibitem{Herzberg:2013:CSD}
A.~Herzberg, H.~Shulman, J.~Ullrich, and E.~Weippl.
\newblock Cloudoscopy: Services discovery and topology mapping.
\newblock In {\em 2013 ACM Workshop on Cloud Computing Security Workshop},
  pages 113--122, 2013.

\bibitem{sandy-bridge-llc}
D.~Kanter.
\newblock L3 cache and ring interconnect.
\newblock \url{http://www.realworldtech.com/sandy-bridge/8/}.

\bibitem{kohno2005remote}
T.~Kohno, A.~Broido, and K.~Claffy.
\newblock Remote physical device fingerprinting.
\newblock In {\em Security and Privacy, 2005 IEEE Symposium on}, pages
  211--225. IEEE, 2005.

\bibitem{httperf}
D.~Mosberger and T.~Jin.
\newblock httperf—a tool for measuring web server performance.
\newblock {\em ACM SIGMETRICS Performance Evaluation Review}, 26(3):31--37,
  1998.

\bibitem{rist:hey-you:ccs:2009}
T.~Ristenpart, E.~Tromer, H.~Shacham, and S.~Savage.
\newblock Hey, you, get off of my cloud: Exploring information leakage in
  third-party compute clouds.
\newblock In {\em Proceedings of the 16th ACM conference on Computer and
  communications security}, pages 199--212. ACM, 2009.

\bibitem{venkat:rfa:ccs2012}
V.~Varadarajan, T.~Kooburat, B.~Farley, T.~Ristenpart, and M.~M. Swift.
\newblock Resource-freeing attacks: Improve your cloud performance (at your
  neighbor's expense).
\newblock In {\em Proceedings of the 2012 ACM conference on Computer and
  communications security}, pages 281--292. ACM, 2012.

\bibitem{venkat-placement15}
V.~Varadarajan, Y.~Zhang, T.~Ristenpart, and M.~Swift.
\newblock A placement vulnerability study in multi-tenant public clouds.
\newblock In {\em 24th USENIX Security Symposium}. USENIX Association, 2015.

\bibitem{who-was-paper}
L.~Wang, A.~Nappa, J.~Caballero, T.~Ristenpart, and A.~Akella.
\newblock Whowas: A platform for measuring web deployments on {IaaS} clouds.
\newblock In {\em Proceedings of the 2014 Conference on Internet Measurement
  Conference}, pages 101--114. ACM, 2014.

\bibitem{wu2012whispers}
Z.~Wu, Z.~Xu, and H.~Wang.
\newblock Whispers in the hyper-space: High-speed covert channel attacks in the
  cloud.
\newblock In {\em USENIX Security symposium}, pages 159--173, 2012.

\bibitem{xu2011exploration}
Y.~Xu, M.~Bailey, F.~Jahanian, K.~Joshi, M.~Hiltunen, and R.~Schlichting.
\newblock An exploration of {L2} cache covert channels in virtualized
  environments.
\newblock In {\em Proceedings of the 3rd ACM workshop on Cloud computing
  security workshop}, pages 29--40. ACM, 2011.

\bibitem{XuWaWu:2015}
Z.~Xu, H.~Wang, and Z.~Wu.
\newblock A measurement study on co-residence threat inside the cloud.
\newblock In {\em USENIX Security Symposium}, 2015.

\bibitem{yarom14flush}
Y.~Yarom and K.~Falkner.
\newblock Flush+reload: A high resolution, low noise, {L3} cache side-channel
  attack.
\newblock In {\em 23rd USENIX Security Symposium}, pages 719--732. USENIX
  Association, 2014.

\bibitem{zhang11:homealone}
Y.~Zhang, A.~Juels, A.~Oprea, and M.~K. Reiter.
\newblock Homealone: Co-residency detection in the cloud via side-channel
  analysis.
\newblock In {\em Proceedings of the 2011 IEEE Symposium on Security and
  Privacy}, pages 313--328. IEEE Computer Society, 2011.

\bibitem{zhang12}
Y.~Zhang, A.~Juels, M.~K. Reiter, and T.~Ristenpart.
\newblock {Cross-VM} side channels and their use to extract private keys.
\newblock In {\em Proceedings of the 2012 ACM conference on Computer and
  communications security}, pages 305--316. ACM, 2012.

\bibitem{zhang2014cross}
Y.~Zhang, A.~Juels, M.~K. Reiter, and T.~Ristenpart.
\newblock Cross-tenant side-channel attacks in {PaaS} clouds.
\newblock In {\em Proceedings of the 2014 ACM SIGSAC Conference on Computer and
  Communications Security}, pages 990--1003. ACM, 2014.

\bibitem{xen-cycle-stealing}
F.~Zhou, M.~Goel, P.~Desnoyers, and R.~Sundaram.
\newblock Scheduler vulnerabilities and attacks in cloud computing.
\newblock {\em CoRR}, abs/1103.0759, 2011.

\end{thebibliography}




\end{document}